%% file: PaperBSHF.tex
\documentclass[%
 reprint,
 superscriptaddress,
nofootinbib,
 amsmath,amssymb,
 aps,
 prd,
 floatfix
]{revtex4-2}

\usepackage{graphicx}
\usepackage{dcolumn}
\usepackage{bm}
\usepackage{amssymb}
\usepackage{color}
\usepackage{siunitx}
\usepackage{hyperref}

\newcommand{\F}{\mathcal{F}}
\newcommand{\Fab}{\F_{\!_{\mathrm{AB}}}}

\newcommand{\vrr}{\vec{r}}
\newcommand{\vv}{\vec{v}}
\newcommand{\va}{\vec{a}}
\newcommand{\vn}{\hat{n}}
\newcommand{\kdot}[1]{^{(#1)}}

\newcommand{\Ord}[1]{\mathcal{O}\left(#1\right)}

\newcommand{\Om}{\Omega}

\newcommand{\SH}{\textsc{SkyHough}}
\newcommand{\BSHF}{\textsc{BinarySkyHou}$\F$}
\newcommand{\BSH}{\textsc{BinarySkyHough}}
\newcommand{\TwoSpect}{\textsc{TwoSpect}}
\newcommand{\coh}{_{\mathrm{c}}}
\newcommand{\sco}{_{\mathrm{s}}}
\newcommand{\seg}{_{\mathrm{seg}}}
\newcommand{\Tseg}{T\seg}
\newcommand{\tref}{\tau_{\mathrm{ref}}}
\newcommand{\fk}[1]{f^{(#1)'}}
\newcommand{\tSSB}{t_{_\mathrm{SSB}}}
\newcommand{\asini}{a_{\mathrm{p}}}
\newcommand{\argp}{\omega}
\newcommand{\tperi}{t_{\mathrm{p}}}
\newcommand{\tasc}{t_{\mathrm{asc}}}
\newcommand{\ecc}{e}
\newcommand{\Porb}{P_\mathrm{orb}}
\newcommand{\mmid}{_{\mathrm{m}}}
\newcommand{\tmid}{t\mmid}
\newcommand{\Ndet}{N_{\mathrm{det}}}
\newcommand{\demod}{_{\mathrm{d}}}
\newcommand{\rec}{_\mathrm{r}}
\newcommand{\avg}[1]{\left\langle #1\right\rangle}
\newcommand{\kmax}{{k_{\max}}}
\newcommand{\iseg}{\ell}

\graphicspath{ {./figs/} {./} }

\usepackage{soul}

\begin{document}

\preprint{APS/123-QED}

\title{Improved all-sky search method for continuous gravitational waves from
  unknown neutron stars in binary systems}

\author{P.~B.~Covas}
\affiliation{Max Planck Institute for Gravitational Physics (Albert Einstein Institute), D-30167 Hannover, Germany}
\affiliation{Leibniz Universität Hannover, D-30167 Hannover, Germany}

\author{R.~Prix}
\affiliation{Max Planck Institute for Gravitational Physics (Albert Einstein Institute), D-30167 Hannover, Germany}
\affiliation{Leibniz Universität Hannover, D-30167 Hannover, Germany}

\input{git_tag.tex}

\date{\commitDATE; \commitIDshort-\commitSTATUS}

\begin{abstract}
  Continuous gravitational waves from spinning deformed neutron stars have not been detected yet,
  and are one of the most promising signals for future detection.
  All-sky searches for continuous gravitational waves from unknown neutron stars in binary systems
  are the most computationally challenging search type.
  Consequently, very few search algorithms and implementations exist for these sources, and only a
  handful of such searches have been performed so far.
  In this paper, we present a new all-sky binary search method, \BSHF{}, which extends and improves
  upon the earlier \BSH{} method, and which was the basis for a recent search
  [\citet{O3aBinary}].
  We compare the sensitivity and computational cost to previous methods, showing that it is both
  more sensitive and computationally efficient, which allows for broader and more sensitive
  searches.
\end{abstract}

\maketitle

\section{\label{sec:intr}Introduction}

Continuous gravitational waves (CWs) are long-lasting and nearly monochromatic gravitational waves,
expected to be emitted by deformed spinning neutron stars (NSs) due to their time-varying
quadrupole moment \citep[e.g.][]{Sieniawska_2019}.
Although many CW searches have been performed to date, using data from the LIGO (H1 and L1) and Virgo (V1) detectors,
no detection has been achieved yet (see \cite{KeithReview} for a recent review).
The expected CW amplitudes are several orders of magnitude smaller than the
compact-binary-coalescence signals currently being routinely detected.
Therefore, the combined analysis of months to years worth of data is required to accumulate enough
signal-to-noise ratio.

When searching for CWs from known pulsars, all the phase-evolution parameters are known from
electromagnetic observations, which allows one to perform statistically optimal searches by coherent
matched-filtering with very little required computing power.
All-sky CW searches for unknown neutron stars represent the opposite extreme, where no prior
information about the signals is available, requiring an expensive explicit search over the
phase-evolution parameters.

Furthermore, because the required parameter-space resolution increases rapidly with longer coherent
integration time, the resulting computing cost explodes and makes it impossible to analyze longer
stretches of data by coherent matched filtering.
This computing-cost problem is pushed to the extreme when searching for unknown neutron stars in
binary systems, as now we also need to search over the unknown binary orbital parameters
\cite{2015arXiv150200914L}.
Therefore all-sky searches for unknown neutron stars in binary systems are the most computationally
challenging type of searches.

The primary strategy followed by computationally-limited CW searches is to break up the data into
shorter segments that can be coherently analyzed individually and then combine these coherent
results across segments.
These are the so-called \emph{semi-coherent} methods (see \cite{universe7120474} for a recent review
of search methods).
The resulting coarser parameter-space resolution entails a reduced computational cost, which allows
for analyzing larger datasets and thereby regaining sensitivity.
As a result, semi-coherent methods are typically more sensitive than fully-coherent matched
filtering at a fixed computational budget \cite{PhysRevD.61.082001,Prix_Shaltev_2012}.

The most commonly used coherent detection statistics are the $\F$-statistic and the Fourier power
(for sufficiently short segments, where a simple sinusoid can approximate the signal).
The $\F$-statistic is obtained by analytically maximizing the likelihood ratio over the four unknown
amplitude parameters of a CW signal \cite{jks98:_data,cutler05:_gen_fstat}.
Although this statistic was initially thought to be optimal, its implicit amplitude
priors have been shown to be unphysical \cite{2009CQGra..26t4013P}.
Using better physical priors results in a more sensitive Bayes factor, albeit (currently) at
increased computational cost, which is why this is not yet a viable alternative to the
$\F$-statistic for wide parameter-space searches.
However, for short segments compared to a day, a new detection statistic has recently been found
\cite{2DOF} that is more sensitive than $\F$ at no extra computing cost, termed the
\emph{dominant-response} statistic $\Fab$.

Using Fourier power over short segments directly as a coherent detection statistic is
computationally cheaper, given that no phase demodulation or other additional calculations are
needed.
However, one limitation of this statistic is the constrained maximum coherent length (about
$\Tseg\lesssim \SI{30}{minutes}$), resulting from the approximation of the signal as a simple
sinusoid. Typically this is expressed as the criterion that the signal power remains in a single
frequency bin (of size $1/\Tseg$).
Furthermore, while demodulated statistics (such as $\F$ and $\Fab$) can naturally combine data from
several detectors coherently \cite{cutler05:_gen_fstat,2DOF}, this is not straightforward to achieve
for short Fourier transforms \cite{Goetz_Riles_2016} and is not commonly used.
Therefore, constructing semi-coherent statistics on demodulated coherent statistics is generally
more sensitive and flexible than using Fourier power.

Only two previous all-sky binary pipelines have been used in searches before
\BSHF{}~\cite{O3aBinary}, namely \TwoSpect{}~\cite{2011CQGra..28u5006G}, and
\BSH{}~\cite{2019PhRvD..99l4019C}.
\TwoSpect{} was the first pipeline for all-sky CW searches of unknown NSs in binary systems
\cite{2014PhRvD..90f2010A}.
\BSH{} is an extension of \SH{}~\cite{PhysRevD.70.082001} (an all-sky pipeline for isolated systems)
to searches for NSs in binary systems, which yields higher sensitivity compared to \TwoSpect{} thanks to its
more sensitive detection statistic and usage of GPU parallelization.
Two recent all-sky binary searches deployed \BSH{} on data from Advanced LIGO's O2 and O3 observing
runs \cite{O2AllSkyBinary,O3aAllSkyBinary}, although over a reduced parameter space in frequency and
binary parameters compared to the \TwoSpect{} search.

\BSH{} uses short-segment Fourier power as its coherent detection statistic, limiting its attainable
sensitivity (as discussed above).
Here we present \BSHF{}, an extension of \BSH{}, which features several improvements compared to the
previous pipeline:
\begin{itemize}
\item Use of demodulated coherent statistics (such as $\F$- or $\Fab$-statistic) instead of
  short-segment Fourier power.
\item Directly summing coherent detection statistics (the typical StackSlide approach
  \cite{PhysRevD.61.082001,Mendell_Landry}) instead of (thresholded) 1s and 0s as in the classical
  Hough algorithm \cite{PhysRevD.70.082001}.
\item Various code-implementation improvements (such as GPU coalesced memory access) and
  optimizations, increasing computational efficiency.
\end{itemize}
As will be shown in this paper, the new search pipeline is both more sensitive and more
computationally efficient than \BSH{}, i.e., for the same coherent segment length $\Tseg$ and
mismatch distribution, it achieves higher sensitivity at lower computational cost.

A key ingredient of the new pipeline is the use of (low-order) Taylor-expanded phase parameters to
describe the binary motion over the (short) coherent segments instead of the physical binary orbital
parameters.
These \emph{Taylor coordinates} allow for a substantial dimensional reduction and solve the problem
of covering the highly-degenerate per-segment coherent parameter space with an efficient template
bank.
However, this approach limits the sensitivity to signals from binary systems with orbital periods
substantially longer than the segment length $\Tseg$.

The development of more sensitive all-sky binary search methods is of utmost importance since more
than half of all known millisecond pulsars are part of a binary
system\footnote{http://www.atnf.csiro.au/research/pulsar/psrcat} \cite{2005AJ....129.1993M}.
Furthermore, accretion from a companion gives a plausible mechanism to generate an asymmetry or
excite an r-mode with a detectable amplitude in the current generation of gravitational-wave
detectors, as recently discussed in \cite{O3aBinary}.

This paper is organized as follows: in Sec.~\ref{sec:phase} we introduce the approximate signal
model used to compute the $\F$-statistic; in Sec.~\ref{sec:upgrades} we present the new \BSHF{}
pipeline and we compare it to its predecessor; in Sec.~\ref{sec:sens-param-estim} we show
sensitivity comparisons of different detection statistics; in Sec.~\ref{sec:conclusions} we
summarize the main results of this paper and lay out some ideas for future work.

\section{\label{sec:phase}Signal phase model}

\subsection{Physical phase model}
\label{sec:physical-phase-model}

Assuming a slowly varying NS spin frequency, the phase of a CW signal in the source frame can be
expressed in terms of the Taylor expansion around a reference time $\tref$, namely
\begin{equation}
  \phi(\tau) = \phi_0 + 2\pi \sum_{k=0}^{s} \frac{f_k}{(k+1)!} (\tau - \tref)^{k+1},
  \label{eq:phaseNS}
\end{equation}
where $\tau$ denotes the time in the source frame, and $s$ is the order of spindown parameters
$f_k$ needed to accurately describe the intrinsic frequency evolution.
The evolution of frequency $f(\tau)$ and higher-order spindowns $\fk{k}(\tau)$ is given by
\begin{equation}
  \label{eq:1}
  f(\tau) = \frac{1}{2\pi} \frac{d \phi(\tau)}{d\tau}, \quad\text{and}\quad
  \fk{k}(\tau) = \frac{d^k f(\tau)}{d\tau^k}\,,
\end{equation}
and the frequency and spindown \emph{parameters} $f_k$ in the phase model of Eq.~\eqref{eq:phaseNS}
are defined at the reference time $\tref$, i.e.,
\begin{equation}
  \label{eq:10}
  f_k \equiv \fk{k}(\tau=\tref)\,.
\end{equation}
In order to obtain the phase of the signal in the frame of a detector, we need to transform it
from the source frame by taking into account the movement of the NS and the movement of the detector
with respect to the solar system barycenter (SSB).
We absorb the unknown relative distance of the source with respect to the SSB into the reference
time $\tref$, and here we neglect relativistic effects such as Shapiro and Einstein
delays\footnote{The numerically-implemented phase model includes these effects for the solar system
  but not for the binary system.} and the transverse proper motion of the source.
We can obtain the timing relation in two steps, first linking the wavefront-emission time $\tau$ in
the source frame to its arrival time $\tSSB$ in the SSB frame, namely
\begin{align}
  \tau(\tSSB) = \tSSB - R(\tau),
  \label{eq:sourceSSB}
\end{align}
where $R(\tau)$ is the radial distance (in light-travel time) of the source to the binary
barycenter (BB) \cite{2015arXiv150200914L}, with $R>0$ when the source is further away from us than
the BB.
In the second step we can relate the SSB time to the arrival time $t$ at detector $X$ by the
Rømer-delay expression:
\begin{equation}
  \label{eq:2}
  \tSSB^X(t) = t + \vrr^X(t)\cdot\vn,
\end{equation}
where $\vrr^X(t)$ is the position vector (in light-travel time) of detector $X$ with respect to the
SSB, and $\vn$ is the sky-position unit vector pointing from the SSB to the BB.
The radial distance $R$ of the source to the BB can be expressed~\cite{2015arXiv150200914L} as
\begin{equation}
  R(\tau) = \asini \left[ \sin{\argp} (\cos{E}-\ecc) + \cos{\argp} \sin{E} \sqrt{1-\ecc^2} \right],
  \label{eq:5}
\end{equation}
in terms of the eccentric anomaly $E$, given by Kepler's equation
\begin{equation}
  E = \Om\,(\tau - \tperi) + \ecc\sin{E},
  \label{eq:8}
\end{equation}
where $\asini$ is the projected semi-major axis of the orbital ellipse (in light-travel time),
$\Om = 2\pi/\Porb$ is the (average) orbital angular velocity (corresponding to the period $\Porb$),
$\ecc$ is the orbital eccentricity, $\tperi$ is the time of periapse passage and $\argp$ is the
(angular) argument of periapse.

For small-eccentricity orbits, this can be approximated by the (linear in $\ecc$) ELL1
model~\cite{Lange_Camilo_Wex_Kramer_Backer_Lyne_Doroshenko_2001,2011arXiv1109.0501M,2015arXiv150200914L}, namely
\begin{equation}
  \label{eq:14}
  R(\tau) = \asini\left[\sin\Psi + \frac{\kappa}{2}\sin{2\Psi} - \frac{\eta}{2}\cos{2\Psi}\right] + \Ord{\ecc^2},
\end{equation}
(dropping a constant term $-3\asini\eta/2$) with the Laplace-Lagrange parameters
defined as $\kappa\equiv e\cos\argp$ and $\eta\equiv e\sin\argp$ and the orbital phase
\begin{equation}
  \label{eq:9}
  \Psi(\tau) = \Om\,(\tau - \tasc)\,,
\end{equation}
using the time of ascending node $\tasc$ instead of the time of periapse passage
$\tperi$, which (to lowest order in $\ecc$) are related by $\tasc = \tperi -\argp/\Om$.
Expressions for $R$ up to any order in
$e$ are found in Appendix~C of \cite{Nieder_Allen_Clark_Pletsch_2020}.

\subsection{Short-segment SSB Taylor coordinates $\{u_k\}$}
\label{sec:short-segment-ssb}

Given all-sky binary CW searches need to cover a huge signal parameter space with finite computing
resources, the longest coherent segment lengths $\Tseg$ that can be used are typically very short
(i.e, much shorter than a day).
If we further assume the orbital periods to be much longer than the short segments, i.e.,
$\Tseg\ll\Porb$, then in this short-segment regime we can resort to Taylor-expanding the phase (in
the SSB) around each segment mid-time $\tmid$ (translated to the SSB), similar to what was done in
\cite{2011arXiv1109.0501M,2015arXiv150200914L}, namely
\begin{equation}
  \label{eq:3}
  \phi(\tSSB) = \phi_0 + 2\pi \sum_{k=1}^\kmax \frac{u_k}{k!}\, (\tSSB - \tmid)^k\,,
\end{equation}
which defines the (SSB) \emph{Taylor coordinates} $\{u_k\}_{k=1}^\kmax$ as
\begin{equation}
  \label{eq:4}
  u_k \equiv \frac{1}{2\pi}\, \left.\frac{d^k\phi}{d\,t_{\mathrm{SSB}}^k}\right|_{\tmid}\,.
\end{equation}
Note that for segments short compared to a day one could also define Taylor coordinates in the
detector frame instead of in the SSB, but this would result in detector-dependent
coordinates that are not suitable for our present search method. The resulting expressions are
given in App.~\ref{sec:appendix} for reference.

Inserting the physical phase model of Eq.~\eqref{eq:phaseNS} in the form
$\phi(\tSSB)=\phi\left(\tau(\tSSB)\right)$, we obtain the phase derivatives\footnote{The general
  form of these successive chain- and product-rule expansions is governed by the so-called Faà di
  Bruno's formula \cite{enwiki:1071944003}.}.
\begin{equation}
  \label{eq:6}
  \begin{aligned}
    u_1 &= \left.\left[ f(\tau) \, \dot{\tau}\right]\right|_{\tmid}\,,\\
    u_2 &= \left.\left[ f'(\tau)\,\dot{\tau}^2 + f(\tau)\,\ddot{\tau} \right]\right|_{\tmid}\,,\\
    u_3 &= \left.\left[ f''(\tau)\,\dot{\tau}^3 + 3f'(\tau)\,\ddot{\tau}\dot{\tau} + f(\tau)\,\dddot{\tau}\right]\right|_{\tmid},\\
    \vdots
  \end{aligned}
\end{equation}
where the derivatives $\tau\kdot{k} \equiv d^k\tau/d\tSSB^k$ of the source-to-SSB timing relation
$\tau(\tSSB)$ can be further expanded using Eq.~\eqref{eq:sourceSSB}, involving derivatives
$R^{(k)}\equiv d^kR(\tau)/d\tSSB^k$ of the binary radial distance $R(\tau)$ of Eq.~\eqref{eq:5},
which can be expanded in the same form as
\begin{equation}
  \label{eq:11}
  \begin{aligned}
    \dot{R} &= R'\,\dot{\tau},\\
    \ddot{R} &= R''\,\dot{\tau}^2 + R'\,\ddot{\tau},\\
    \dddot{R} &= R'''\,\dot{\tau}^3 + 3R''\,\ddot{\tau}\dot{\tau} + R'\,\dddot{\tau},\\
    \vdots
  \end{aligned}
\end{equation}
in terms of the source-frame time derivatives $R^{(k)'}\equiv d^kR(\tau)/d\tau^k$.
This analysis is complicated by the fact that the binary radial distance $R(\tau)$ of Eq.~\eqref{eq:5} is
a function of source-frame (emission) time $\tau$, not the SSB (arrival) time $\tSSB$ of a wavefront.
In \cite{2011arXiv1109.0501M,2015arXiv150200914L} this difficulty could be neglected for the
purposes of computing the parameter-space metric, where a slow-orbit approximation, i.e.,
$R(\tau)\approx R(\tSSB)$, is sufficient.  However, in the present application we want to preserve
higher accuracy for the purpose of using these coordinates for coherent matched-filtering.

Substituting into the timing derivatives of Eq.~\eqref{eq:sourceSSB}, we can now obtain the
expressions
\begin{equation}
  \label{eq:18}
  \begin{aligned}
    \dot{\tau} &= \left[1+R'\right]^{-1} , \\
    \ddot{\tau} &= \left[1+R'\right]^{-1}\left(-R''\,\dot{\tau}^2\right),\\
    \dddot{\tau} &= \left[1+R'\right]^{-1}\left(-R''\dot{\tau}^3 - 3R''\,\ddot{\tau}\dot{\tau}\right),\\
    \vdots
  \end{aligned}
\end{equation}
which are explicit because of the iterative backwards dependency of the $\tau^{(k)}$ on only
lower-order derivatives, i.e.,
$\tau^{(k)} = \tau^{(k)}(\dot{\tau},\ddot{\tau},\ldots \tau^{(k-1)})$.

From these expressions we can obtain the explicit Taylor coordinates $u_k$ via Eq.~\eqref{eq:6} as
\begin{align}
  u_1 &= \frac{f\mmid}{1+R'\mmid},   \label{eq:u1}\\
  u_2 &= -\frac{f\mmid\,R''\mmid}{\left(1+R'\mmid\right)^3} + \frac{f'\mmid}{\left(1 + R'\mmid\right)^2},  \label{eq:u2}
\end{align}
where for the present application (such as \cite{O3aBinary}) at most first- or second-order
$u_k$ will be needed in practice, as discussed in Sec.~\ref{sec:CohMis}.
Here we defined
\begin{equation}
  \label{eq:13}
  \begin{aligned}
    f\mmid &\equiv f(\tmid) = f_0 + f_1\,(\tmid - \tref) + \ldots,\\
    f'\mmid &\equiv f'(\tmid) = f_1 + f_2\,(\tmid - \tref) + \ldots,
  \end{aligned}
\end{equation}
and
\begin{equation}
  \label{eq:16}
  \begin{aligned}
    R'\mmid &= \asini\Om\left[\cos\Psi\mmid + \kappa\cos2\Psi\mmid + \eta\sin2\Psi\mmid\right],\\
    R''\mmid &= -\asini\Om^2\left[\sin\Psi\mmid + 2\kappa\sin2\Psi\mmid - 2\eta\cos2\Psi\mmid\right],
  \end{aligned}
\end{equation}
where $R^{(k)'}\mmid \equiv R^{(k)'}(\tmid)$ and
\begin{equation}
  \label{eq:20}
  \Psi\mmid \equiv \Om\left(\tmid - \tasc\right),
\end{equation}
assuming the small-eccentricity approximation of Eq.~\eqref{eq:14}, and (here) neglecting the NS-BB
time delay of Eq.~\eqref{eq:sourceSSB} as a higher-order correction, i.e.,
$\tau(\tmid)\approx\tmid$.

These $u_1,u_2$ coordinates have units of \si{Hz} and \si{Hz^2}, respectively, and depend on the
physical parameters $\{\{f_k\},\asini,\Om, \tasc, \ecc, \argp\}$.
Using these coordinates assumes that we have performed the standard SSB demodulation of the signal
for any given sky position $\vn$, which is typically expressed in terms of the right ascension
$\alpha$ and declination $\delta$ in equatorial coordinates.

The resulting (constant) parameter-space metric for the Taylor coordinates $\{u_k\}$ (valid for
any signal phase of the form Eq.~\eqref{eq:3}) is found in Eq.~(57) of \cite{2015arXiv150200914L}.

\section{\BSHF{}}
\label{sec:upgrades}

In this section we present a summary of the new \BSHF{} pipeline and its main advantages over the previous \BSH{}.

\subsection{Summary of the previous and new pipeline}
\label{sec:summary-previous-new}

The predecessor \SH{} and \BSH{} algorithms are described in more detail in \cite{PhysRevD.70.082001} and
\cite{2019PhRvD..99l4019C}, here we only provide a short overview summary.
Both of these analyze the frequency-time matrix of short-Fourier-transform (SFT) power, by searching for
``tracks'' (corresponding to different source parameters) that are above the statistical expectation
for noise.

\SH{} is limited to searches for signals from isolated systems, while \BSH{} is an extension
designed for all-sky searches for unknown neutron stars in binary systems.
Both are extremely fast model-based pipelines due to the highly efficient algorithms used to analyze
the sky-maps and their effective use of look-up tables (see
\cite{PhysRevD.70.082001,2019PhRvD..99l4019C} for details).
Furthermore, \BSH{} leverages the computational advantages provided by GPUs by parallelizing the
most expensive steps in the algorithm, and thus further massively reducing the runtime of a search.

A \BSH{} search is divided in two consecutive stages, using different detection statistics. In the
first stage, a ``Hough'' weighted sum of 1s and 0s (depending on whether the SFT power crossed a
threshold or not) is calculated, and all of the templates are sorted by the resulting significance
(a normalized Hough number count with normal distribution, see Eq.~(25) of \cite{2019PhRvD..99l4019C}).
The frequency-time pattern used for the tracks in the first stage is an approximation to the exact
one, due to the usage of look-up tables (explained in Sec.~IV B of
\cite{2019PhRvD..99l4019C}).
In the second stage, the \emph{refinement stage}, a fraction of the best-ranked templates is
analyzed again, this time using a StackSlide weighted sum of SFT power, which has a higher
sensitivity than summing weighted 1s and 0s (e.g., see \cite{Mendell_Landry}), and using a more accurate expression for the frequency-time pattern.

The central new feature of the \BSHF{} pipeline is to use a \emph{demodulated} coherent detection
statistic for the segments\footnote{The $\F$-statistic has been used before in combination with the
  Hough algorithm, in \cite{Aasi_2013}, an all-sky search for isolated systems using day-long
  coherent segments, where the (single-stage) pipeline summed weighted 1s and 0s computed from
  thresholded $\F$-statistic values.}, such as the $\F$- \cite{jks98:_data,cutler05:_gen_fstat} or
$\Fab$-statistic \cite{2DOF}, instead of the number count or SFT power, but otherwise still benefit
from the highly-efficient GPU-based \BSH{}-type algorithm to combine the coherent results to a
semi-coherent statistic.
Three main benefits arise from using a demodulated coherent statistic:
\begin{enumerate}
  \item Demodulation removes the constraint on the maximum segment length, as the signal is no longer
approximated as a pure sinusoid. This allows the algorithm to turn increases in computing power into
better sensitivity (shown in Sec.~\ref{sec:SensTc}).
  \item The per-detector data is combined coherently, which reduces the number of coherent
segments needed to combine in the semi-coherent stage, improving sensitivity (shown in
Sec.~\ref{sec:Sens1}) and reducing computational cost (shown in Sec.~\ref{sec:CompCost}).
  \item The parameter-space resolution and resulting mismatch can be controlled as a free parameter
(rather than the fixed $1/\Tseg$ Fourier resolution of SFTs), which will be discussed in
Sec.~\ref{sec:CohMis}.
\end{enumerate}

When combining coherent results to calculate a semi-coherent detection statistic, it has been shown
that applying per-segment weights can improve the sensitivity \cite{Krishnan_Sintes}.
While there is currently no analytic argument for using a weighted sum of $\F$- or
$\Fab$-statistics\footnote{This is left for future work, but intuitively can be understood from the
  per-segment change in signal power, which can vary by around one order of magnitude between
  segments for such short coherent times.}, empirically we find (shown in Sec.~\ref{sec:Sens1}) that
using weights also improves the sensitivity for these detection statistics.
The weight $w_\iseg$ at segment $\iseg$ is given by:
\begin{align}
  w_\iseg &= K \frac{A_\iseg + B_\iseg}{S_{\mathrm{n;}\iseg}} ,\\
  K &= \frac{N\seg}{\sum_{\iseg=1}^{N\seg}w_\iseg} , \\
  A_\iseg &= \sum_{N_{\mathrm{SFTs}}} a^2, \quad B_\iseg = \sum_{N_{\mathrm{SFTs}}} b^2,
\end{align}
where $K$ is a normalization factor, $S_{\mathrm{n;}\iseg}$ is the noise power spectral density of segment $\iseg$, and $a$ and $b$ are the antenna patterns of a detector (evaluated at the mid-time of every SFT), where the sum goes over all the SFTs in segment $\iseg$. When the segment just has one SFT, we recover the weights given by Eq.~(22) of \cite{2019PhRvD..99l4019C}. For the dominant-response statistic, we use the following weights:
\begin{align}
  w_\iseg = K \begin{cases} \frac{A_\iseg + \frac{C_\iseg^2}{A_\iseg}}{S_{\mathrm{n;}\iseg}} & \text{if } A_\iseg \ge B_\iseg, \\
  \frac{B_\iseg + \frac{C_\iseg^2}{B_\iseg}}{S_{\mathrm{n;}\iseg}} & \text{otherwise},
  \end{cases}
\end{align}
where $C_\iseg = \sum_{N_{\mathrm{SFTs}}} a\,b$.

As discussed in Sec.~\ref{sec:short-segment-ssb}, for computing-cost reasons the coherent segments
for all-sky binary searches need to be very short, which allows us to use a small number (currently
one or two) of Taylor coordinates $u_k$ to represent the spindown and binary orbital motion in the
coherent segments.  Using physical parameters, we would need to build a (at least) 6-dimensional
parameter space grid\footnote{In comparison to \BSH{} the new code is able to also search over
  spindowns and eccentricity in the semi-coherent stage.}  over
$\{\alpha,\delta,f_0, \asini,\Om,\tasc\}$, while using the Taylor coordinates we effectively only
need to use three (or four) parameter-space dimensions for the short segments currently considered, namely
$\{\alpha,\delta, u_1(, u_2)\}$.  This reduces complexity (the parameter-space metric in physical
parameters would be hugely degenerate for short segments
(cf.~\cite{2011arXiv1109.0501M,2015arXiv150200914L}) and lowers the resulting computational
cost\footnote{Using a Taylor phase approximation to lower the computational cost of an all-sky
  binary search has been first explored by the \textit{Polynomial} pipeline
  \cite{Putten_Bulten_Brand_Holtrop_2010}, which did not use physical parameters for the
  semi-coherent summation, however, resulting in reduced sensitivity.}.

The $u_k$ template bank is constructed as a hyper-cubic lattice in coordinate space.  The code
processes the sky in patches defined by an isotropic grid with cells of fixed solid-angle, using
partial Hough map derivatives \cite{2019PhRvD..99l4019C} to process the semi-coherent sky mapping.
Coherent per-segment statistics are computed only for the center of each sky-patch using the
corresponding antenna pattern modulations and weights.

\subsection{Semi-coherent interpolation}
\label{sec:semi-coher-interp}

In the previous section we obtained the Taylor coordinates $u_k$, which together with the sky
position coordinates will be used in the coherent stage in order to calculate the
$\mathcal{F}$-statistic values over coordinates $\{\alpha,\delta, \{u_k\}\}$.
In the semi-coherent stage, on the other hand, we are using physical coordinates to combine the
per-segment statistics, namely $\{\alpha,\delta,\{f_k\}, \asini, \Om, \tasc, \ecc, \argp\}$.
For every semi-coherent template, we therefore need to calculate the appropriate \emph{mapping} to
the corresponding Taylor coordinates $\{u_k\}$ and coherent sky position.

In addition to using different signal parameters, the semi-coherent template grids also generally
need to be finer than the coherent per-segment grids, which results in the need to
\emph{interpolate} the coherent results when combining them semi-coherently (typically using
nearest-neighbor interpolation).
This is a generic feature of the semi-coherent approach (cf.\
\cite{Prix_Shaltev_2012,wette_implementing_2018}), and in \SH{}-derived pipelines takes the form of
the so-called \emph{master equation} \cite{PhysRevD.70.082001,2019PhRvD..99l4019C}, linking sky
offsets to resulting effective frequency shifts of the signal.

The \SH-type sky interpolation works by breaking the sky into several sky patches, as mentioned
above, where the center of each patch is used as the coherent sky template for every semi-coherent
sky-template in the same sky patch.
The resulting offset in sky-position between the semi-coherent and coherent template results in
compensating offsets in the $\{u_k\}$ coordinates, generalizing the Hough master equation.

A simple way to derive the shift in $u_k$ coordinates due to an offset $\delta \vn$ in sky
position is to use the full detector-frame Taylor coordinates $u_k^X$ for each detector $X$, given in the appendix in Eq.~\eqref{eq:17}.
Using this we can express the induced shifts $\delta u^X_k$ as
\begin{equation}
  \label{eq:19}
  \begin{aligned}
    \delta u^X_1 &= u_1\,\vv\mmid^X\cdot\delta\vn,\\
    \delta u^X_2 &\approx u_1\,\va\mmid^X\cdot\delta\vn + 2\vv\mmid^X\cdot\delta\vn\left(-u_1\,R\mmid'' + f\mmid'\right)\,,
  \end{aligned}
\end{equation}
in terms of detector velocity $\vv^X\mmid$ and acceleration $\va\mmid^X$ at the segment mid-time
$\tmid$ translated to the SSB\footnote{Strictly speaking, here it should be the segment mid-time in
  the detector frame, but the maximal shift of $\sim\SI{500}{\second}$ can be neglected for detector
  velocity and acceleration.}.  To remove the detector dependency we average over detectors, which
will be a good approximation for $\delta u_1$, given the detector velocity is dominated by the
(detector-independent) orbital motion of the Earth.  On the other hand, the detector acceleration
$\va\mmid^X$ in $\delta u_2^X$ is dominated by the Earth's rotation, so averaging over detectors
might be a less reliable approximation, but should still work well as long as the detectors are not
too far separated, such as for LIGO H1 and L1.
Therefore we arrive at the following generalized master equations
\begin{align}
  u_{\mathrm{I}1} &= u_1\,\left( 1 + \vv\mmid\cdot\delta\vn\right),   \label{eq:21a}\\
  u_{\mathrm{I}2} &= u_2 + u_1\,\va\mmid\cdot\delta\vn + 2\vv\mmid\cdot\delta\vn\left(-u_1\,R\mmid'' +
           f\mmid'\right),            \label{eq:21b}
\end{align}
with detector-averaged velocity $\Ndet\,\vv \equiv \sum_X \vv^X$ and acceleration
$\Ndet\,\va = \sum_X \va^X$.
Eq.~\eqref{eq:21a} agrees with the previous Hough-on-$\F$-statistic master equation found in
\cite{PhysRevD.70.082001} and \cite{Aasi_2013} (with implicit detector averaging).

\begin{figure}
  \centering
  \includegraphics[width=1.0\columnwidth]{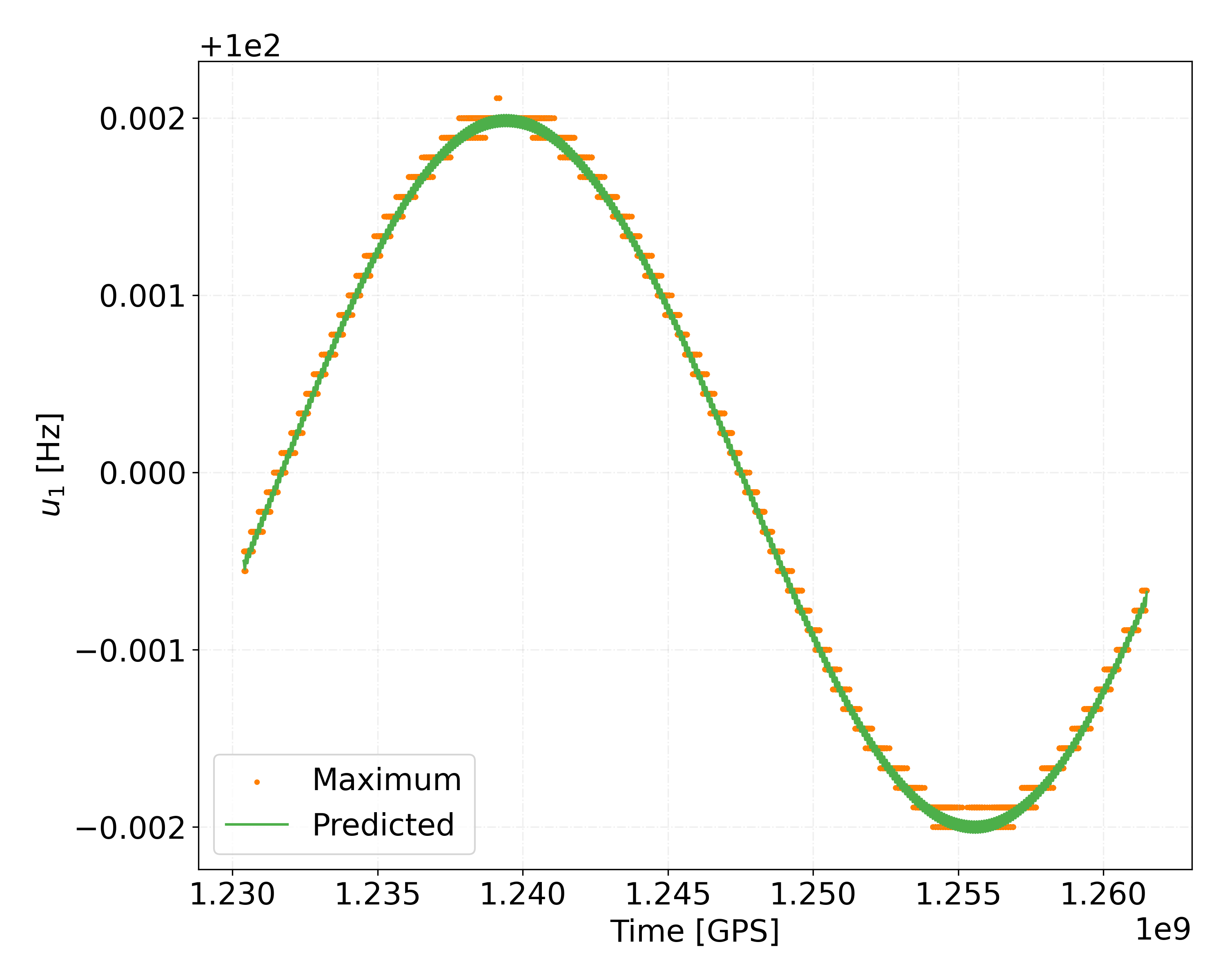}
  \caption{Comparison between the $u_{\mathrm{I}1}$ coordinate per segment given by
    Eq.~\eqref{eq:21a} (green line) and the $u_1$ values per segment that maximize the signal power
    (orange points). The sky position is offset by \SI{0.3}{rad} in both $\alpha$ and
    $\delta$ from the signal. This example assumes a single detector (H1) and one year of data, with
    segments of length $\Tseg = \SI{900}{\second}$, and a constant-frequency signal of
    $f_0=\SI{100}{Hz}$ without binary modulation.}
  \label{fig:U1Shift}
\end{figure}

The $u_{\mathrm{I}1}$ master equation is illustrated in Fig.~\ref{fig:U1Shift}, where the $u_1$ values with the highest signal power are
plotted as a function of segment mid-time $\tmid$ for an offset $\delta\vn = \vn-\vn\demod$ between the
signal sky-position $\vn$ and the coherent demodulation point $\vn\demod$. In addition we plot the predicted track of shifted $u_{\mathrm{I}1}$ given by Eq.~\eqref{eq:21a}. These $u_{\mathrm{I}1}$ values closely follow the path that minimizes the mismatch.
The mismatch in Fig.~\ref{fig:U1Shift} between the path followed by $u_{\mathrm{I}1}$ and the maximum path
is around $0.1 \%$, whereas the mismatch between the non-shifted $u_{1}$ and the maximum path would
be around $\SI{80}{\percent}$.

The parameter-space bounds for each of the Taylor coordinates are found using Eqs.~\eqref{eq:21a} and
\eqref{eq:21b}, by calculating the maximum possible values over the given physical parameter space.

For computational efficiency reasons (namely the look-up table approach used here, see
\cite{PhysRevD.70.082001}) the first-stage ``Hough'' semi-coherent summation actually uses the
following approximate expressions instead of Eqs.~\eqref{eq:21a} and \eqref{eq:21b}, namely
\begin{align}
  u_{\mathrm{IH}1} &= f\mmid (1 - R'\mmid) + f_{\mathrm{H}}\, \vv\mmid\cdot\delta\vn, \\
  u_{\mathrm{IH}2} &= u_2,
\end{align}
with the ``middle'' frequency $f_{\mathrm{H}}$ given by
\begin{align}
 f_{\mathrm{H}} \equiv f_{0\mathrm{H}} + f_{1\mathrm{H}} (t_{\mathrm{mid}} - \tref) + \ldots,
\end{align}
where the $f_{k\mathrm{H}}$ denote the midpoints in frequency- and spindown-ranges currently being
searched over, and $t_{\mathrm{mid}}$ is the mid-time of the full dataset.

\subsection{Mismatch}

In this subsection we describe and characterize different sources of mismatch for the \BSHF{}
pipeline.
The mismatch is defined as the relative loss of signal power, namely
\begin{align}
  \mu = 1 - \frac{\rho\rec^2}{\rho^2},
\end{align}
where $\rho^2$ is the full signal power (given by Eq.~20 in \cite{2DOF}) and $\rho\rec^2$ is the signal power recovered by the search.

The total mismatch of the \BSHF{} pipeline has several contributions, which can be separated in
coherent and semi-coherent mismatches. The main contributions to the coherent mismatch are offsets
between signal and the closest template in the coherent template grid, and the usage of the
(truncated) Taylor coordinates $u_k$, while the semi-coherent mismatch is produced by
signal-template offsets in the semi-coherent grid and approximations in the interpolation mapping
(discussed in the previous section).

\subsubsection{Mismatch due to Taylor-coordinate truncation}
\label{sec:CohMis}

The usage of a limited number of Taylor coordinates $u_k$ incurs an intrinsic mismatch due to the
corresponding approximation of the signal waveform.
In practice we currently envisage using only $u_1$ or at most up to order $u_2$, which turns out to
be sufficient for currently considered practical all-sky binary searches (similar to the recent
search \cite{O3aBinary} using only $u_1$) due to computational constraints.
Therefore we quantify the mismatch and corresponding constraints on the maximum coherent segment
length $\Tseg$ due to truncation to order $u_1$ or $u_2$.

The mismatch $\mu_{u_k}$ due to omission of order $u_k$ (and higher) can be estimated as
$\mu_{u_k} \sim g_{kk}\,u_k^2$ using the metric $g_{kl}$ in $u_k$-space, which is given in
Eqs.~(56, 57) of \cite{2015arXiv150200914L}.
Using this we can express the mismatch due to truncation of $u_{k\ge2}$ or $u_{k\ge3}$ as
\begin{equation}
  \label{eq:23}
  \begin{aligned}
    \mu_{u_2} &\approx g_{22}\,u_2^2 = \frac{\pi^2\,\Tseg^4}{180}\,u_2^2,\\
    \mu_{u_3} &\approx g_{33}\,u_3^2 = \frac{\pi^2\,\Tseg^6}{4032}\,u_3^2.
  \end{aligned}
\end{equation}
Using a time-average $\avg{.}$ over segments together with Eq.~\eqref{eq:u2} for $u_2$ we obtain
$\avg{u_2^2}\approx \frac{1}{2}f_0^2\asini^2\Om^4 + f_1^2$ (neglecting smaller corrections), and for
$u_3$ we can use Eq.~(58) of \cite{2015arXiv150200914L} as an estimate, which yields
$\avg{u_3^2}\approx \frac{1}{2}f_0^2\asini^2\,\Om^6$, and so we obtain the (segment-averaged) mismatch
estimates as
\begin{align}
  \avg{\mu_{u_2}} &\approx \frac{\pi^2\,\Tseg^4}{360}\,\left( f_0^2\asini^2\Om^4 + 2f_1^2\right), \label{eq:mismU1av}\\
  \avg{\mu_{u_3}} &\approx \frac{\pi^2\,\Tseg^6}{8064}\,f_0^2\asini^2\,\Om^6\,,
  \label{eq:mismU2av}
\end{align}
which illustrate the fact that the segments must be short compared to the orbital period, i.e.,
$\Tseg\,\Om\ll 1$, in order for the Taylor-coordinates $u_k$ to be a good approximation, as
discussed in \cite{2015arXiv150200914L}.

\begin{figure*}
  \centering
  \includegraphics[width=1.0\columnwidth]{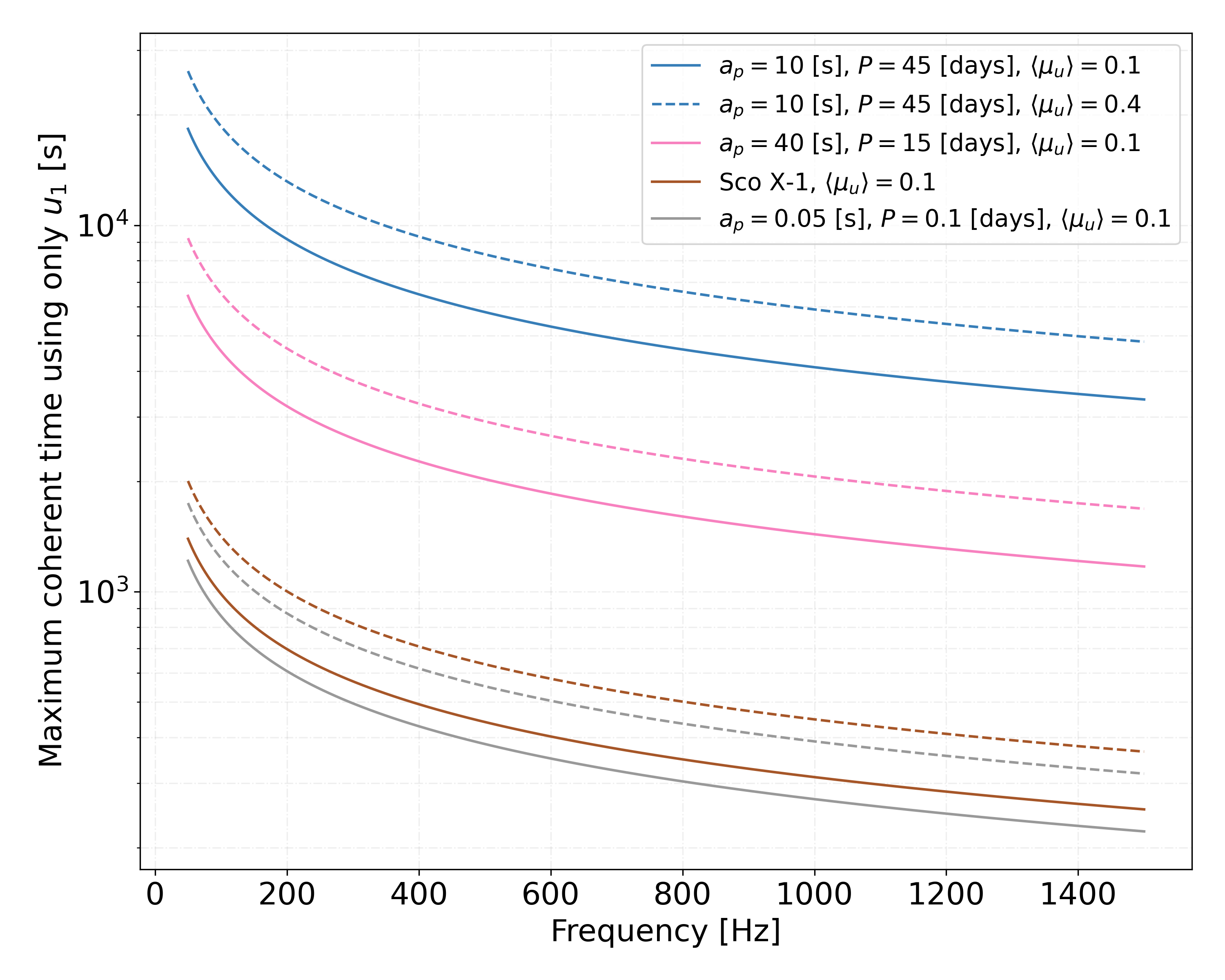}
  \includegraphics[width=1.0\columnwidth]{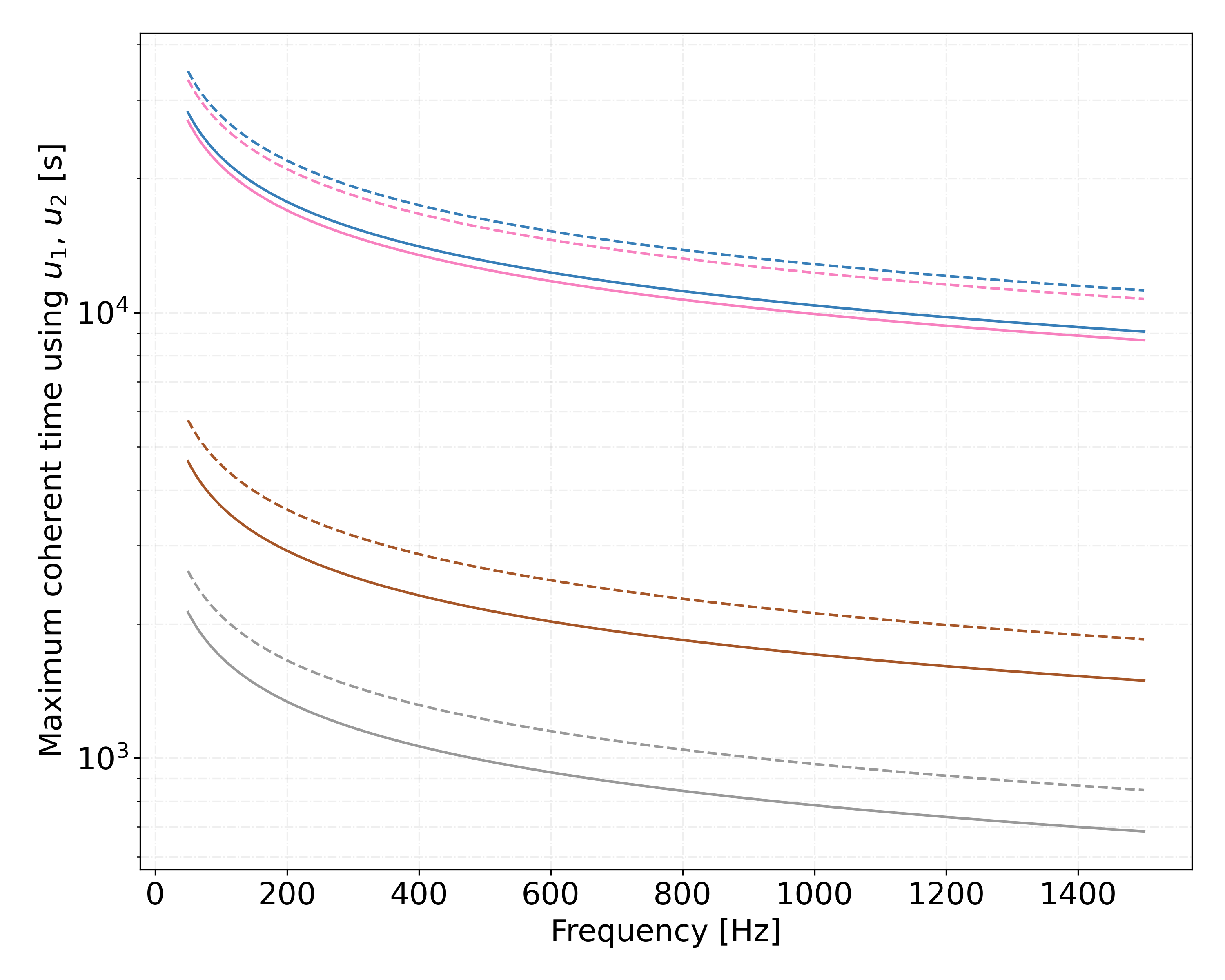}
  \caption{The left plot shows the maximum coherent time $\Tseg$ that can be used in a given region of the
    binary parameter space with a certain mismatch due to the Taylor truncation (assuming zero
    spindown $f_1=0$).
    The left plot shows the results for $\avg{mu_u}=0.1$ (solid lines) and $\avg{\mu_u}=0.4$ (dashed
    lines) in Eq.~\eqref{eq:maxTc}, using only $u_1$ Taylor coordinate.
    The right plot shows the same when using both $u_1$ and $u_2$ Taylor coordinates.
    }
  \label{fig:MaxTc}
\end{figure*}

We can rearrange these equations to obtain a constraint on the maximum coherent time $\Tseg$
allowed for a given acceptable mismatch $\avg{\mu_u}$ from Taylor truncation, namely
when using only $u_1$ we find
\begin{align}
  \Tseg{}_{,u_1} &\leq  \left( \frac{360 \,\avg{\mu_u}}{\pi^2 \left( f_0^2 \asini^2 \Om^4 + 2 f_1^2 \right)} \right)^{1/4},
  \label{eq:maxTc}
\end{align}
and similarly for truncation after $u_2$ we obtain the constraint
\begin{align}
  \Tseg{}_{,u_2} &\leq \left( \frac{ 8064 \,\avg{\mu_u}}{\pi^2 f_0^2 \asini^2 \Om^6} \right)^{1/6}.
  \label{eq:maxTc2}
\end{align}
These expressions for the maximum coherent time are illustrated in Fig.~\ref{fig:MaxTc} as a
function of frequency for different choices of binary orbital parameters. It can be seen that when $u_2$ is also used the maximum coherent time increases by a certain factor.

\begin{figure}
  \centering
  \includegraphics[width=1.0\columnwidth]{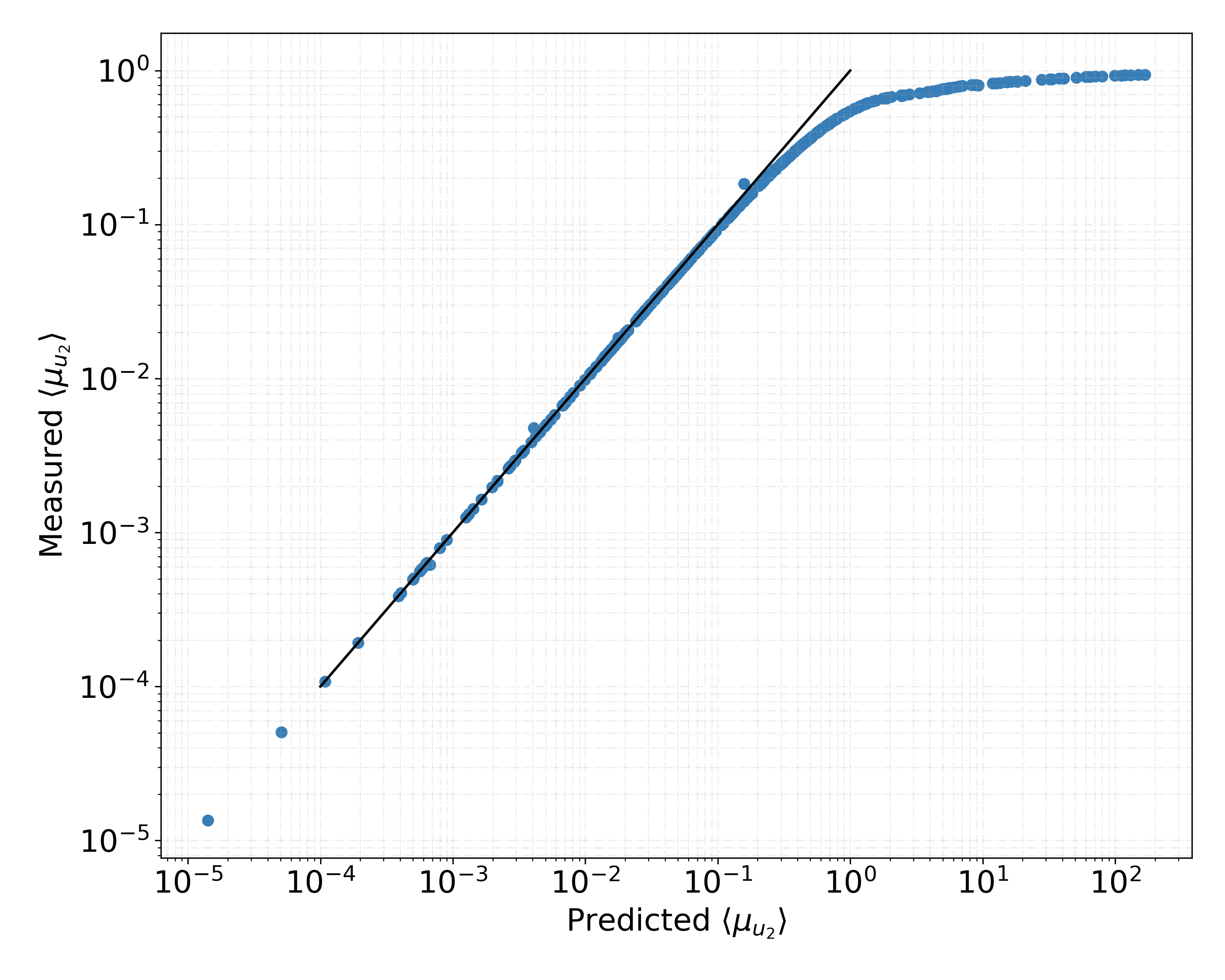}
  \caption{Measured mismatch $\avg{\mu_{u_2}}$ (blue points) from Taylor truncation (averaged over
    segments) against the predicted mismatch (solid line) given by Eq.~\eqref{eq:mismU1av}, using
    \num{1000} injections and assuming one year of data from the H1 detector with coherent segments of
    $\Tseg = \SI{900}{\second}$.
  }
  \label{fig:MaxTcInjections}
\end{figure}

Figure~\ref{fig:MaxTcInjections} shows a test of the mismatch predicted by Eq.~\eqref{eq:mismU1av}, by
generating \num{1000} different signals with a frequency of \SI{300}{Hz} and random orbital
parameters log-uniformly distributed, with $\Om\in[0.1,1]\,\si{days}$ and
$\asini\in[0.01,1]\,\si{ls}$.
For each signal we measure the perfectly-matched signal power when using physical coordinates and
compare it to the signal power obtained with Taylor coordinates up to order $u_1$.
The corresponding measured mismatch is then compared to the model prediction of
Eq.~\eqref{eq:mismU1av}.
The figure shows that these equations correctly predict the measured mismatch, in the range where
the metric approximation is expected to be accurate (i.e., below mismatches of $\sim 0.3$).

\subsubsection{Total mismatch}

In the previous subsection we discussed the mismatch contribution due to using a limited set of
Taylor coordinates.
Additionally there will be template-bank mismatches incurred from the coherent
and semi-coherent template grids. If we count the Taylor-truncation mismatch $\mu_u$ as part of the
coherent mismatch $\mu\coh$, then the total average (over the template bank) mismatch $\avg{\mu}$
will be given approximately \cite{Prix_Shaltev_2012} by the sum of the mean coherent $\avg{\mu\coh}$ and mean semi-coherent
mismatch $\avg{\mu\sco}$, namely
\begin{align}
  \left< \mu \right> \simeq \avg{\mu\coh} + \avg{\mu\sco}.
\end{align}
From this expression it can be seen that if the mean coherent mismatch is reduced while the semi-coherent mismatch is equal, the total mismatch would decrease.

\begin{figure}
  \centering
  \includegraphics[width=1.0\columnwidth]{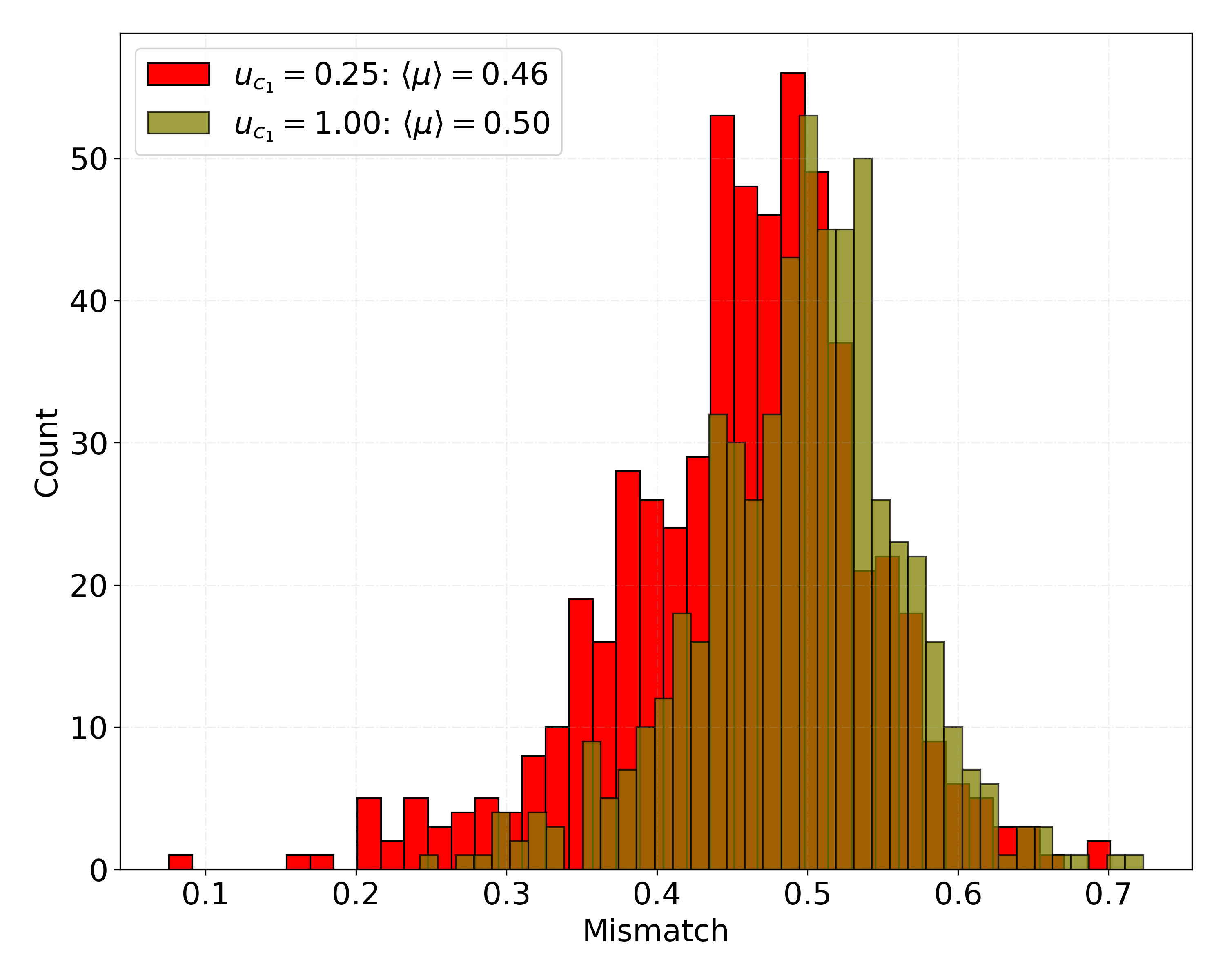}
  \caption{Mismatch histograms for \num{1000} injections with random parameters, assuming one year
    of data from both H1 and L1 detectors and coherent segments of $\Tseg = \SI{900}{\second}$.
    The olive (right) histogram corresponds to a grid resolution of $\delta u_1 = 1/\Tseg$, while the red
    (left) histogram uses a finer resolution of $\delta u_1 = 0.25/\Tseg$.
    Only $u_1$ and a single sky position are searched in the coherent stage. The semi-coherent grids
    are equal for both cases.}
  \label{fig:MismTotal}
\end{figure}

This is shown in Fig.~\ref{fig:MismTotal}, where the total measured mismatch can be seen for two
different cases, which have different $u_1$ coherent grids but the same semi-coherent grid (the
coherent sky position is the same for both cases, and it is shifted from the signal value). A
decrease in the total mismatch can be seen for the case where the coherent $u_1$ grid is finer, as
predicted by the previous equation. This represents an improvement over the previous pipeline, where
the coherent frequency grid was fixed to be equal to the Fourier transform spacing.

\subsubsection{Maximum spindown and eccentricity}
\label{sec:maximum-spindown}

Although \BSHF{} is able to search the $f_1$ spindown and eccentricity $\ecc$ parameters in the
semi-coherent summation, all-sky searches for neutron stars in binary systems are so computationally
expensive that one would usually not explicitly search over these parameters at first. For this
reason, we want to estimate the value ranges in these omitted parameters to which the pipeline is
still sensitive, which will depend on the particular set-up (such as the amount of data and the
coherent time).

The maximum covered spindown value $|f_1|$ is important for interpreting astrophysical upper limits,
since this limits the maximum ``mountain height'' (NS deformation) or r-mode amplitude that
the search can find. The reason for this is that larger deformation (or mode amplitude) would
produce a larger spindown $|f_1|$ due to the emission of gravitational waves, which would
potentially become undetectable by such a search if it was too large.

The semi-coherent grid is constructed by requiring a maximum mismatch $\mu_{\mathrm{M}}$. We can
estimate the maximum allowed values by calculating the required resolution for these parameters:
\begin{equation}
  \label{eq:maxSpEc}
  \begin{aligned}
    |f_{1;\mathrm{max}}| \equiv \delta f_1 &= \sqrt{ \frac{\mu_{\mathrm{M}}}{g_{f_1 f_1}} } = \sqrt{\frac{45 \mu_{\mathrm{M}}}{4 \pi^2 \Tseg^2 T^2_{\mathrm{obs}} }} , \\
    \ecc_{\mathrm{max}} \equiv \delta \ecc &= \sqrt{ \frac{\mu_{\mathrm{M}}}{g_{\ecc\ecc}} } =
    \sqrt{ \frac{6 \mu_{\mathrm{M}}}{\pi^2 \Tseg^2 f_0^2\,\Om^2 \asini^2} },
  \end{aligned}
\end{equation}
where $g_{f_1 f_1}$ is obtained from \cite{Pletsch_2010} assuming that the refinement factor is
equal to $N\seg$ (i.e., there are no gaps), and $g_{\ecc\ecc}$ is found in \cite{2015arXiv150200914L}.

It can be seen that for spindown $f_1$ the maximum value only depends on the segment length $\Tseg$
and total amount of data, while the maximum eccentricity $\ecc$ depends on the frequency $f_0$ and
orbital parameters $\asini$ and $\Om$. To obtain a limit, one can take the parameters that produce the most
conservative eccentricity, or calculate a mean value over the parameter-space boundaries. The
eccentricity equation has the exact same functional form as Eq.~(43) in \cite{2019PhRvD..99l4019C}, while
here we make explicit the dependence on the desired maximum mismatch.

The previous equations quantify the additional mismatch produced by a signal with nonzero
spindown $f_1$ and eccentricity $\ecc$. However, another potential side-effect would be a shift in
the remaining estimated parameters due to correlations between the parameters.
However, values exceeding the limits above would not automatically mean such signals are
undetectable, only that the resulting mismatch would be larger, thus decreasing the sensitivity of the search.

\begin{figure}
  \centering
  \includegraphics[width=1.0\columnwidth]{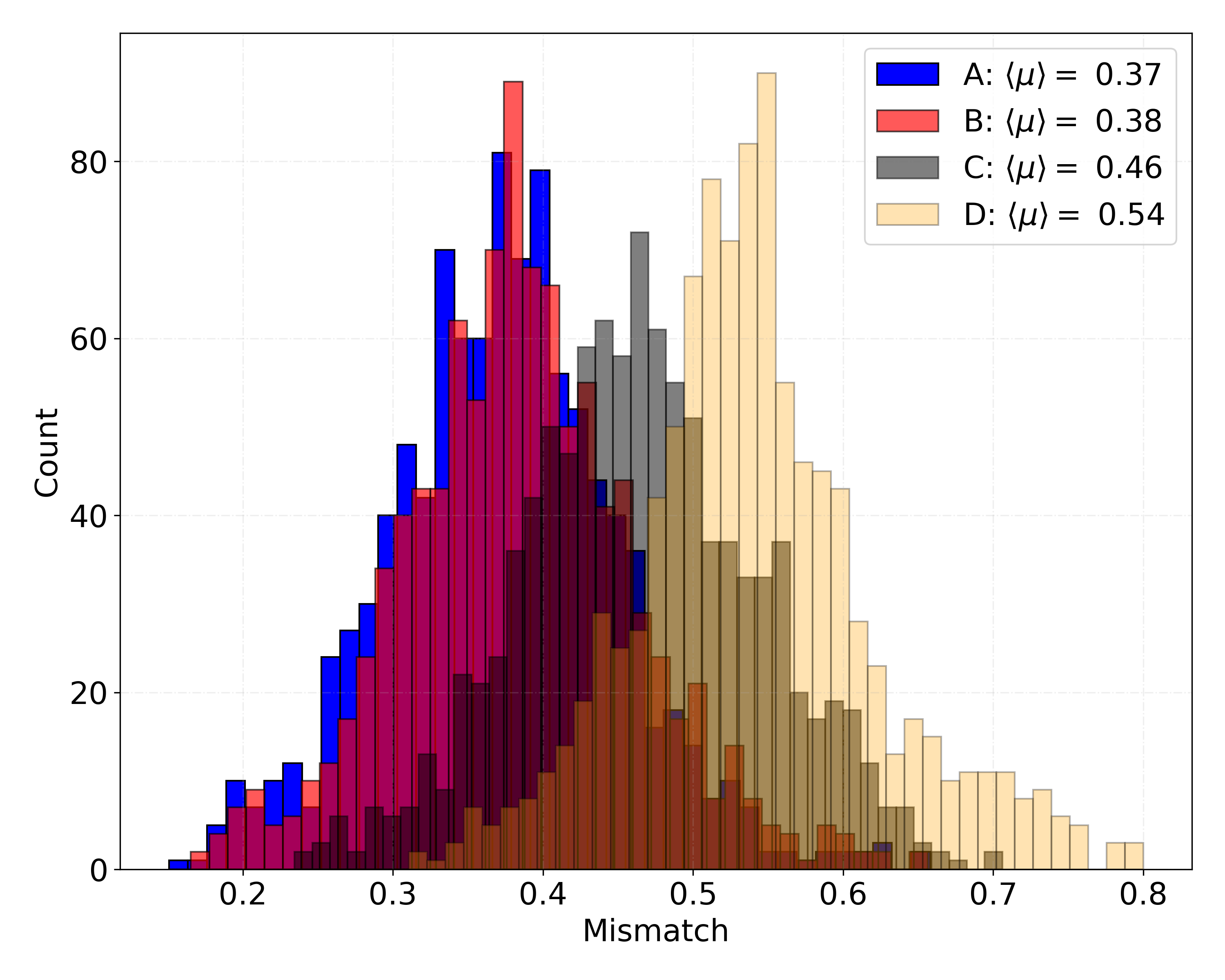}
  \caption{Measured mismatch for \num{1000} injections with random parameters, assuming one year of
    data from the H1 detector and segments of $\Tseg = \SI{900}{\second}$.
    From left to right, the blue (A) histogram corresponds to signals with zero spindown, $f_1=0$,
    and eccentricity $\ecc=0$; the red (B) histogram uses signals
    log-uniformly distributed in
    $f_1\in[-4.7 \times 10^{-14},-4.7 \times 10^{-11}]$~Hz/s and $\ecc\in[0.2 \times 10^{-4},0.2\times 10^{-1}]$;
    the black (C) histogram uses signals with $f_1$ and $e$ at their maximum values given by
    Eq.~\eqref{eq:maxSpEc}; the yellow (D) histogram is for signals with $f_1$ and $e$ given by twice
    the maximum values.}
  \label{fig:MaxSpEcc}
\end{figure}

We can compare the mismatch distribution obtained with the same grid, for four different cases: signals with $f_1 = 0$ and $e = 0$; signals with the maximum values; signals with values in between (with log-uniform distributions up to the maximum value); signals with double the maximum value. This is shown in Fig.~\ref{fig:MaxSpEcc}, where it can be seen that signals with parameters at the maximum value (the eccentricity maximum has been calculated using the binary parameters that give the largest eccentricity) increase the mean mismatch by $\sim 0.08$, which would reduce the sensitivity of a search by $\sim 5\%$.

\subsection{Computational model}
\label{sec:CompCost}

In this section we explain how the computational cost and Random Access Memory (RAM) of our pipeline scale with different set-up variables, updating and expanding Sec.~VF of \cite{2019PhRvD..99l4019C}.

\subsubsection{Coherent computational cost}

Due to the calculation of the $\mathcal{F}$-statistic values, the coherent stage will have an
additional computational cost, besides loading the input data and calculating the partial Hough map
derivatives. In order to estimate this cost, we summarize the content of
\cite{prix:Fstat_timing}. The cost of calculating the $\mathcal{F}$-statistic or its related
quantities in segment $\iseg$ at a single sky-patch scales with\footnote{This assumes the so-called
  \textit{Demod} $\F$-statistic implementation, which is computationally favored for this type of search,
  but the pipeline can just as easily use the \textit{Resampling} implementation \cite{prix:Fstat_timing}.}:
\begin{align}
  \tau_{\mathcal{F};\iseg} &= N_{\mathrm{T;}\iseg} \,( N_{u} N_{\mathrm{dterms}} \tau_{\mathrm{core}} + \tau_{\mathrm{buffer}} ), \\
  N_{\mathrm{T;}\iseg} &= \sum_{X=1}^{\Ndet} N_{\mathrm{SFT;}\iseg,X} ,
\end{align}
where $\tau_{\mathrm{core}}$ and $\tau_{\mathrm{buffer}}$ are fundamental timing constants that only
depend on the hardware and optimization settings (usually $\tau_{\mathrm{buffer}}$ is approximately
one order of magnitude bigger), $2 N_{\mathrm{dterms}} + 1$ are the number of frequency bins that
are used for the calculation of the Dirichlet kernel, $N_{u}$ is the number of coherent $u_k$
templates, and $N_{\mathrm{T;}\iseg}$ is the total number of SFTs in segment $\iseg$.

The total coherent computational cost of a single sky-patch scales with the number of segments:
\begin{align}
  \tau_{\mathcal{F}} = \sum_{\iseg=1}^{N_{\mathrm{seg}}} \tau_{\mathcal{F};\iseg} = N_{\mathrm{SFT}} ( N_{u} N_{\mathrm{dterms}} \tau_{\mathrm{core}} + \tau_{\mathrm{buffer}} ).
\end{align}
Since the calculations for each segment are independent from the others, these steps can be easily parallelized. We use an OpenMP loop to take advantage of multi-core CPUs, which can speed-up the calculation by approximately the number of used cores.

The total coherent computational cost will scale linearly with the number of sky-patches.

\subsubsection{Semi-coherent computational cost}

In the semi-coherent stage the coherent detection statistics are combined for every template that is searched.

The cost of the first stage $\tau_{H;j}$ over a sky-patch $j$ scales as:
\begin{equation}
  \label{eq:24}
  \tau_{\mathrm{H;}j} = N_{\mathrm{fs}} N\seg\, b(N_{\mathrm{binary}})\, g_j(N_{\delta},N_{\alpha},\delta_{u_1}, \delta_s)\, h(r)\, \tau_{1},
\end{equation}
where $N_{\mathrm{fs}}$ is the number of semi-coherent frequency and spindown templates, $b()$ is a function containing the non-linear dependency on the number of binary templates $N_{\mathrm{binary}}$, $g_j()$ is a function describing the effective number of semi-coherent sky points needed (due to the \SH{} algorithm), $N_{\alpha}$ and $N_{\delta}$ are the number of right ascension and declination points in each sky-patch, $\delta_s$ is the semi-coherent sky grid resolution, $h()$ is a function that depends on the threshold $r$ set at the coherent stage, and $\tau_1$ is a fundamental timing constant. The total semi-coherent cost will scale with the number of sky-patches.
In the previous paper \cite{2019PhRvD..99l4019C} it was assumed that $b = N_{\mathrm{binary}}$ and $g_j = N_{\delta} N_{\alpha}$, which left some details out.

The function $b()$ depends on the GPU architecture. If we use a CPU, it would simply be equal to $N_{\mathrm{binary}}$, but if we use a GPU it depends non-linearly on parameters such as the occupation of the GPU cores and the usage of shared memory. This can be seen in Fig.~\ref{fig:Comp1}, where the non-linear scaling with $N_{\mathrm{binary}}$ is clear.

The function $g_j()$ is equal or less than $N_{\delta} N_{\alpha}$, and it encodes the \SH-type sky interpolation mechanism, which depends on the relation between the size of the annulus produced by the Doppler modulations and the size of the semi-coherent sky grid, as explained in Sec.~IVB of \cite{PhysRevD.70.082001}. At a given timestamp, the sky-patches with $\hat{n}$ more parallel to $\vec{v}$ have wider annulus, which may contain several semi-coherent sky pixels, thus lowering the number of sky points that need to be taken into account in the semi-coherent loop. This effect will be different at each timestamp, and over a long observing run this will produce an average value between one and $N_{\delta} N_{\alpha}$ for the function $g_j()$. This effect gives the \SH{} algorithm a computational advantage.

The function $h$ is different from one for a non-zero threshold $r$, which substitutes coherent values to 0 when below the threshold, thus reducing the computational cost. This function is given by $h = e^{- \frac{r}{\left<r\right>}}$, where $\left<r\right>$ is the expected value of the coherent detection statistic.

We define the average cost $\avg{\tau_{\mathrm{H}}}$ of the first stage over different sky-patches $j$:
\begin{align}
  \avg{\tau_{\mathrm{H}}} = N_{\mathrm{fs}} N\seg\, b(N_{\mathrm{binary}}) \avg{g_j(N_{\delta},N_{\alpha},\delta_{u_1}, \delta_s)}\, h(r)\, \tau_{1}.
\end{align}
The cost $\tau_{\mathrm{R}}$ of the second (i.e., ``refinement'') stage scales as:
\begin{align}
  \tau_{\mathrm{R}} = N\seg\, b(N_{\mathrm{cand}}) N_{\mathrm{a}}\, \tau_{2},
\end{align}
where $N_{\mathrm{cand}}$ is the number of candidates that are passed to the second stage, $N_{\mathrm{a}}$ is the number of additional points around each candidate that are searched (when using a finer grid), $b()$ is the same function as before, and $\tau_2$ is a fundamental timing constant.

The total semi-coherent computational cost is the sum of the first and second stage.

\subsubsection{Total computational cost}

In order to estimate the total computational cost of a search, we add the coherent and semi-coherent costs (neglecting other costs such as loading the data and writing output to files, since in a realistic scenario they are negligible):
\begin{align}
  \tau = \sum_{l=1}^{N_{\mathrm{F}}} N_{\mathrm{SP;}l} (\tau_{\mathcal{F}} + \left< \tau_{\mathrm{H}} \right>_l + \tau_{\mathrm{R;}l}),
\end{align}
where $N_{\mathrm{F}}$ is the number of frequency bands needed to cover a certain frequency range, and $N_{\mathrm{SP;}l}$ is the number of sky-patches at frequency band $l$.
The values for $\left< \tau_{\mathrm{H}} \right>_l$ and $\tau_{\mathrm{R;}l}$ depend on the frequency band, since $N_{\mathrm{binary}}$, $N_{\delta}$, $N_{\alpha}$, and $N_{\mathrm{cand}}$ scale with frequency.

\begin{figure}
  \centering
  \includegraphics[width=1.0\columnwidth]{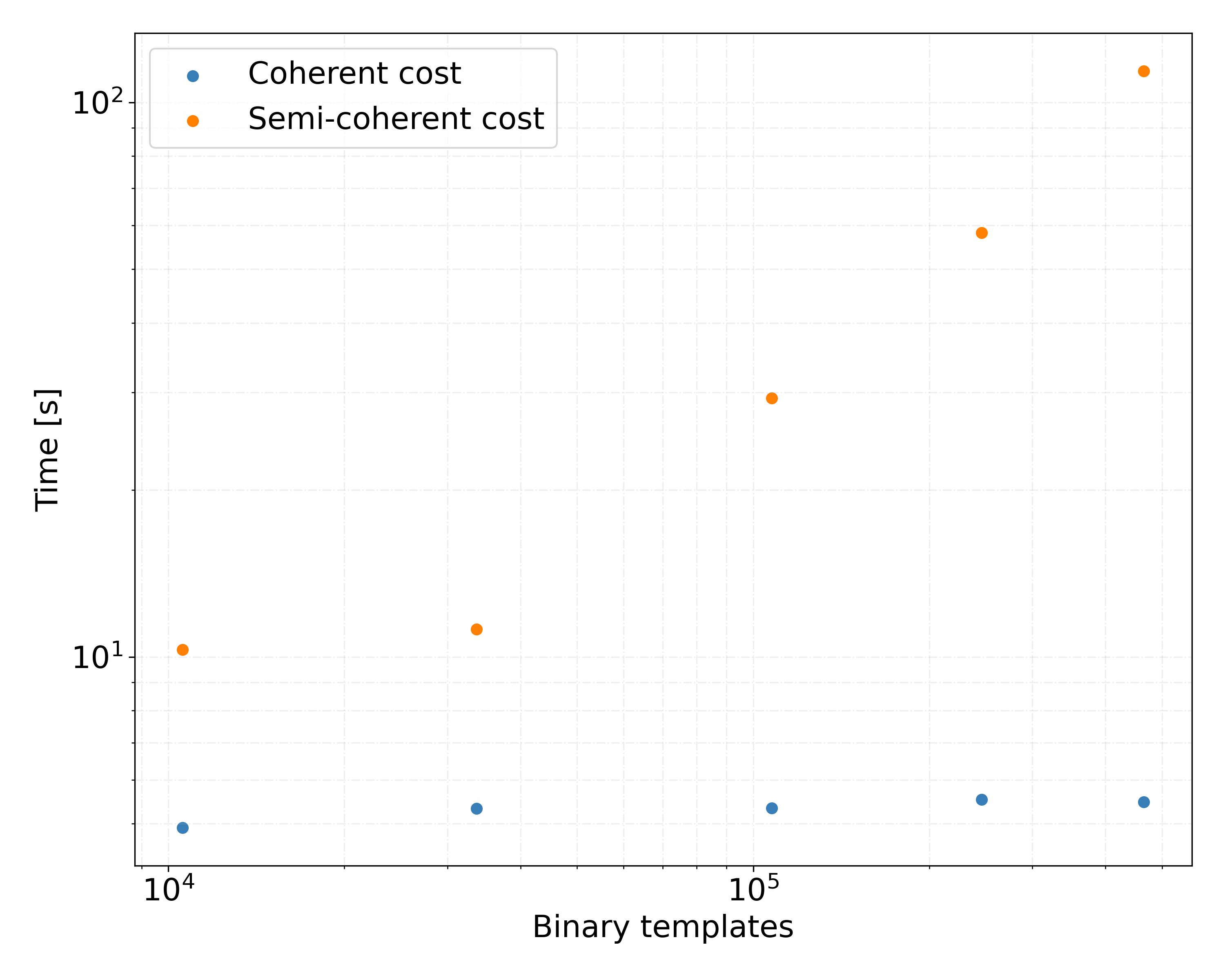}
  \caption{Timing for a single sky-patch of a \SI{0.1}{Hz} frequency band as a function of the
    number of binary templates. The orange points show the computing time of the semi-coherent
    stage, while the blue points are for the coherent stage (including the $\mathcal{F}$-statistic
    computation and generation of partial Hough map derivatives).  These timings are obtained on an
    NVIDIA Tesla V100 and a single core of an Intel(R) Xeon(R) Silver 4215 CPU 2.50 GHz.  }
  \label{fig:Comp2}
\end{figure}

Figure~\ref{fig:Comp2} shows a comparison of the coherent and semi-coherent costs as a function of the number of binary templates. It can be seen that the coherent cost stays constant, but the semi-coherent cost increases, as expected. Past searches using \BSH{} and \BSHF{} have used $N_{\mathrm{binary}}$ larger than $10^5$, so in such typical scenarios the coherent cost will be a small fraction of the total cost.
If the number of binary templates is small, the coherent computational cost will have a non-negligible impact on the total cost. This also happens for isolated searches, where the search over binary parameters is substituted with a search over $f_1$ values, which usually is much less than $10^5$. In these two cases, calculating the $\mathcal{F}$-statistic might lower the sensitivity or the span of a search.

\begin{figure}
  \centering
  \includegraphics[width=1.0\columnwidth]{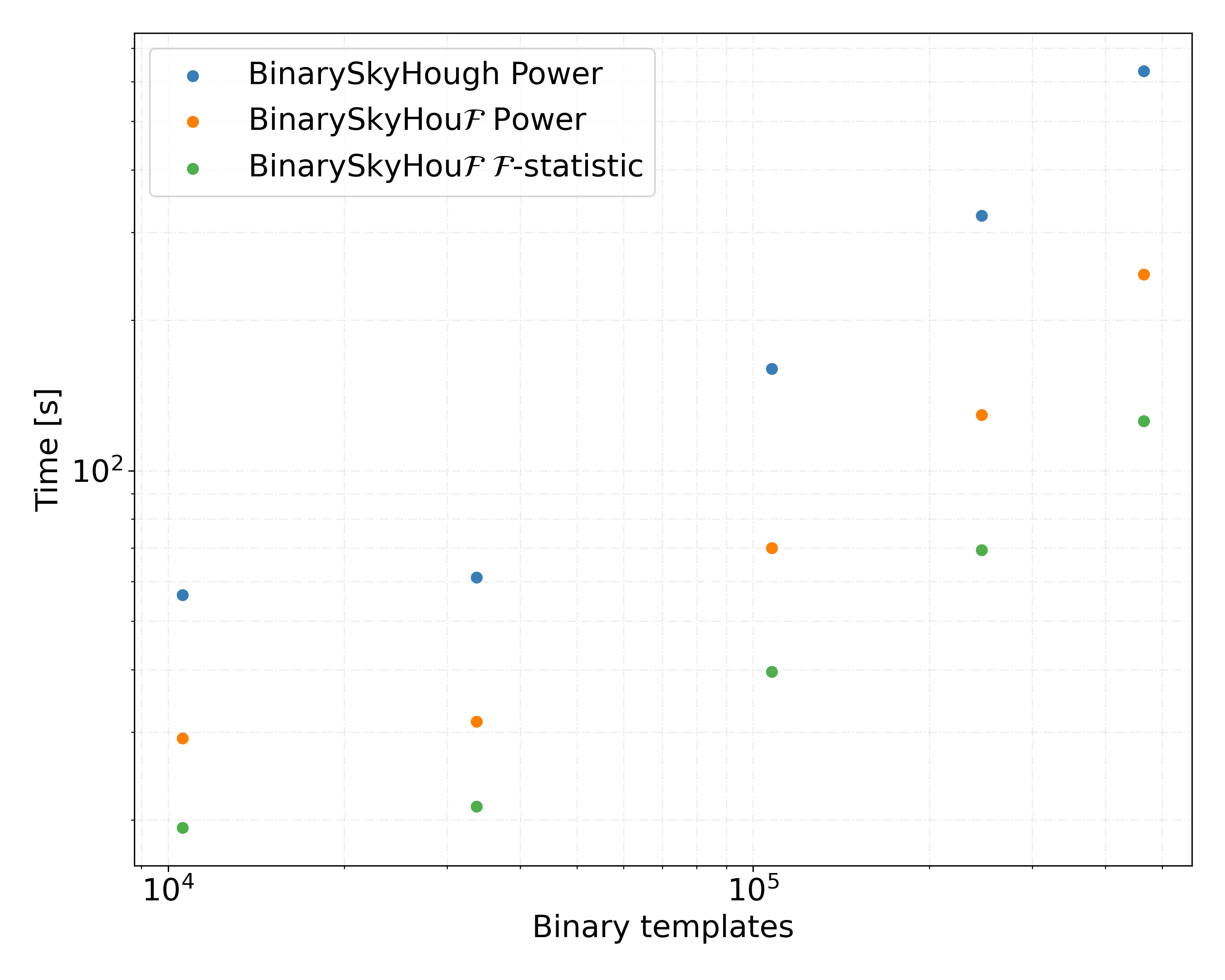}
  \caption{Timing for a single sky-patch of a \SI{0.1}{Hz} frequency band as a function of the
    number of binary templates. The blue points show the cost of the search for the \BSH{} pipeline
    using SFT power as the coherent detection statistic, the orange points show the cost for the new
    \BSHF{} pipeline using the power statistic, while the green points refer to using the
    $\mathcal{F}$-statistic. All three searches use the same number of templates and the same amount
    of data (Gaussian noise with equal amounts of data from detectors H1 and L1).  These timings are
    obtained on an NVIDIA Tesla V100 and a single core of an Intel(R) Xeon(R) Silver 4215 CPU 2.50
    GHz.  }
  \label{fig:Comp1}
\end{figure}

When comparing the computational cost to the previous pipeline, it is important to notice that when
using the $\mathcal{F}$-statistic as the coherent detection statistic, the total computational cost
is reduced compared to using SFT power for multi-detector searches. This is because the
semi-coherent summing of the power is done over $\sim N\seg \Ndet$ values,
while the $\mathcal{F}$-statistic generates just $N\seg$ coherent detection statistics.
For this reason, in the semi-coherent stage the combination of powers will take roughly
$\Ndet$ times longer than the combination of $\mathcal{F}$-statistic values. If the
template grids are the same in both cases, the computational cost of a
semi-coherently dominated search using the $\mathcal{F}$-statistic will therefore be reduced. This improvement
can be seen in Fig.~\ref{fig:Comp1}, where the green points show the lowering of the computational
cost compared to the orange points.

Furthermore, the computational efficiency of the code has been improved, mainly by better usage of
CUDA's coalesced global memory access.
This can be seen in Fig.~\ref{fig:Comp1}, comparing the performance of the previous to the new code
for the same setup as a function of the number of binary templates (increasing the RAM memory
needs).
The increased efficiency of the new code corresponds to a lowering of the timing coefficient $\tau_1$
in Eq.~\eqref{eq:24}, manifesting as a weaker scaling with the number of binary templates.

\subsubsection{RAM}

In order to estimate the RAM required by \BSHF{}, we find the scaling of the data structures in the
code as a function of input parameters such as the maximum mismatch, the coherent time, and the
amount of data used.

The biggest data structures in the code are:
\begin{itemize}
  \item The partial Hough map derivatives, which hold the results from the coherent stage:
  \begin{align}
  S_{\mathrm{P}} = 6 N\seg N_{u} N_{\mathrm{sky}},
  \end{align}
  where $N_{u}$ is the total number of Taylor $u$-coordinate templates, and $N_{\mathrm{sky}} =
  N_{\delta} N_{\alpha}$ is the number of sky templates.
  \item The per-frequency bin semi-coherent results:
  \begin{align}
  S_{\mathrm{R}} = 8 N_{\mathrm{binary}} N_{\mathrm{sky}}.
  \end{align}
\end{itemize}
These structures are orders of magnitude larger than the rest and are therefore enough to give a
good estimate of the required RAM.

The main differences with the previous pipeline are:
\begin{itemize}
  \item The number of frequency bins needed in the partial Hough map derivatives slightly decreases due to the coherent sky demodulation.
  \item The effective number of ``segments'' is reduced by the coherent multi-detector combination.
  \item If more than one Taylor coordinate needs to be used to maintain the mismatch at a certain level, the RAM increases in order to hold the coherent results. This will limit the number of Taylor coordinates that can be used at a certain coherent time.
\end{itemize}

Another RAM limitation of the code is due to the usage of CUDA's shared memory in the GPU kernel functions. This limits the size of the sky-patches, which is dependent on the GPU architecture. The shared memory size is given by:
\begin{align}
  S_{\mathrm{S}} = 4 T N_{\mathrm{sky}},
\end{align}
where $T$ is the number of threads per block in the GPU kernel launch.

\section{Sensitivity and parameter estimation}
\label{sec:sens-param-estim}

In this section we will estimate the sensitivity of the new pipeline, and compare it to the previous
one. In order to do this, we will compare the different detection statistics being used, showing the
improvements in sensitivity that are possible due to the usage of demodulated statistics. We are not
attempting to estimate a realistic sensitivity that these pipelines would achieve in an actual
search, since this also depends on other post-processing procedures, such as clustering or
follow-up, which are beyond the scope of the present study. 

To estimate the sensitivity, we will follow the same common procedure used in
\cite{2019PhRvD..99l4019C}. Namely, for different setups (encompassing the amount of data, detectors
and their relative noise levels, maximum mismatch parameters, and coherent time) and detection
statistics we generate Gaussian noise and perform a search to obtain the threshold at a certain false
alarm probability (in the results shown, we use the top template as the threshold).
Next, we add six groups (each with a different gravitational-wave amplitude) of \num{1000} randomly
distributed signals to the previously generated Gaussian noise, and perform a separate search for
each signal. The resulting statistic values are compared to the threshold, and the detection
probability is estimated by the fraction of signals detected (i.e., crossing that threshold) out of
the total number of signals.
This procedure is performed for every different detection statistic.

The injected signals have a random isotropic (NS spin) orientation, a random isotropic sky position,
a random frequency $f_0$ between $[100,100.1]$ Hz, and a random period $\Porb\in [15,60]$~days and
semi-major axis $\asini \in [10,40]$~ls.

The detection statistics that we compare are the original Hough number count (given by Eq.~(25) of
\cite{2019PhRvD..99l4019C}), the SFT power (given by Eq.~(26) of \cite{2019PhRvD..99l4019C}), the
$\mathcal{F}$-statistic (given by Eq.~(23) of \cite{2DOF}), and the dominant-response statistic (given
by Eq.~(34) of \cite{2DOF}). For each of these statistics we also compare their weighted versions
(discussed in Sec.~\ref{sec:summary-previous-new}.
Here show the results for a single setup as an illustration, but we have tested various setups with
different amounts of data and mismatch distributions and have obtained similar results.

\subsection{Comparison of detection statistics}
\label{sec:Sens1}

\begin{figure*}
\centering
\includegraphics[width=1.0\columnwidth]{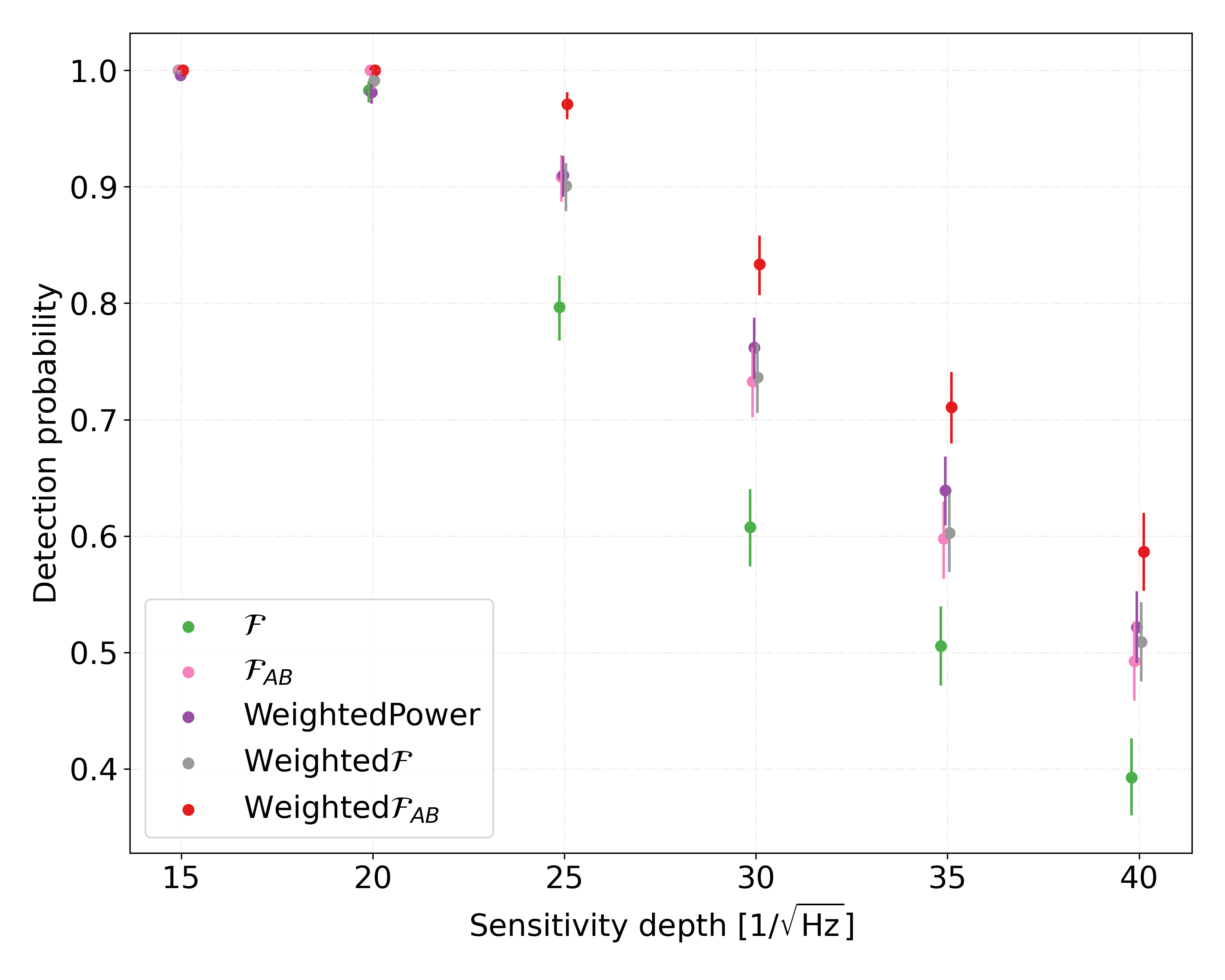}
\includegraphics[width=1.0\columnwidth]{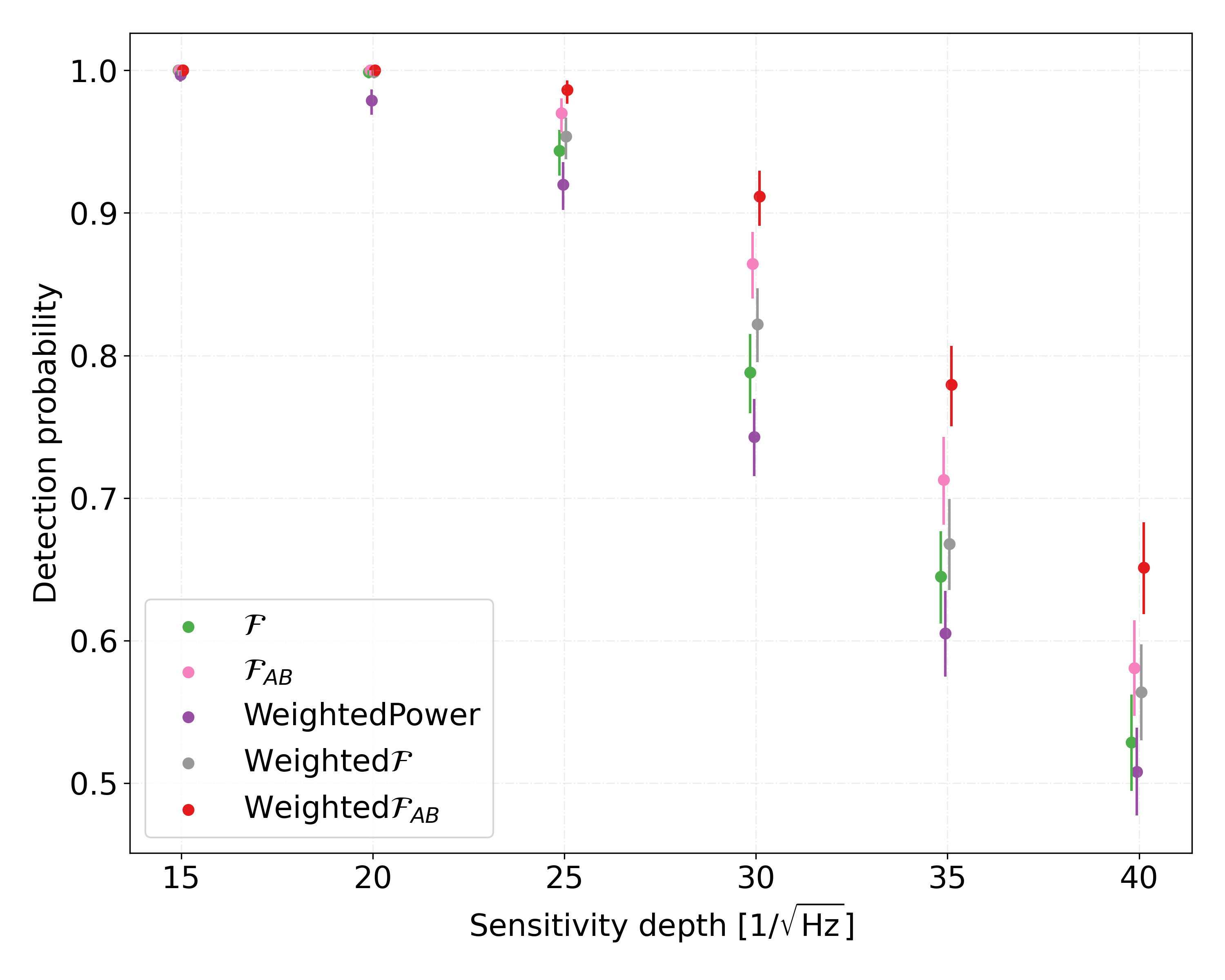}
\caption{Detection probability as a function of sensitivity depth $\sqrt{S_{\mathrm{n}}}/h_0$ for
  different detection statistics in Gaussian noise.  The left plot shows the results for the
  H1 and L1 detectors, while the right plot additionally includes the V1 detector.  The error bars
  denote the $95\%$ binomial confidence interval.  We assume one year of data at the same noise
  floor from each of the detectors, and a coherent segment length of $\Tseg=\SI{900}{\second}$.
  (The points have been slightly displaced along the x-axis to ease visibility).  }
\label{fig:Sens2}
\end{figure*}

The left plot of Fig.~\ref{fig:Sens2} shows the sensitivity of two unweighted ($\F$ and $\Fab$) and
three weighted detection statistics (SFT power, $\F$ and $\Fab$) for two detectors (H1 and L1),
while the right plot shows the same comparison for three detectors (using V1 in addition).
In both plots we see that the most sensitive statistic is the weighted dominant-response
statistic $\Fab$.
For the two-detector case, the sensitivities of the SFT power and $\mathcal{F}$-statistic are within
the statistical errors of each other, but the right plot shows that for more than two detectors the
$\mathcal{F}$-statistic is more sensitive. This is expected from the reduced number of effective
segments (as discussed earlier), resulting in reduced $\chi^2$ degrees of freedom for the background
distribution (e.g., see \cite{2DOF}).
We further see that the sensitivity of all statistics improves (on these short segments) when using
weights. Overall these results illustrate the sensitivity advantage of the demodulated ($\F$- or
$\Fab$-) statistics over using SFT power.

\begin{figure}
\centering
\includegraphics[width=1.0\columnwidth]{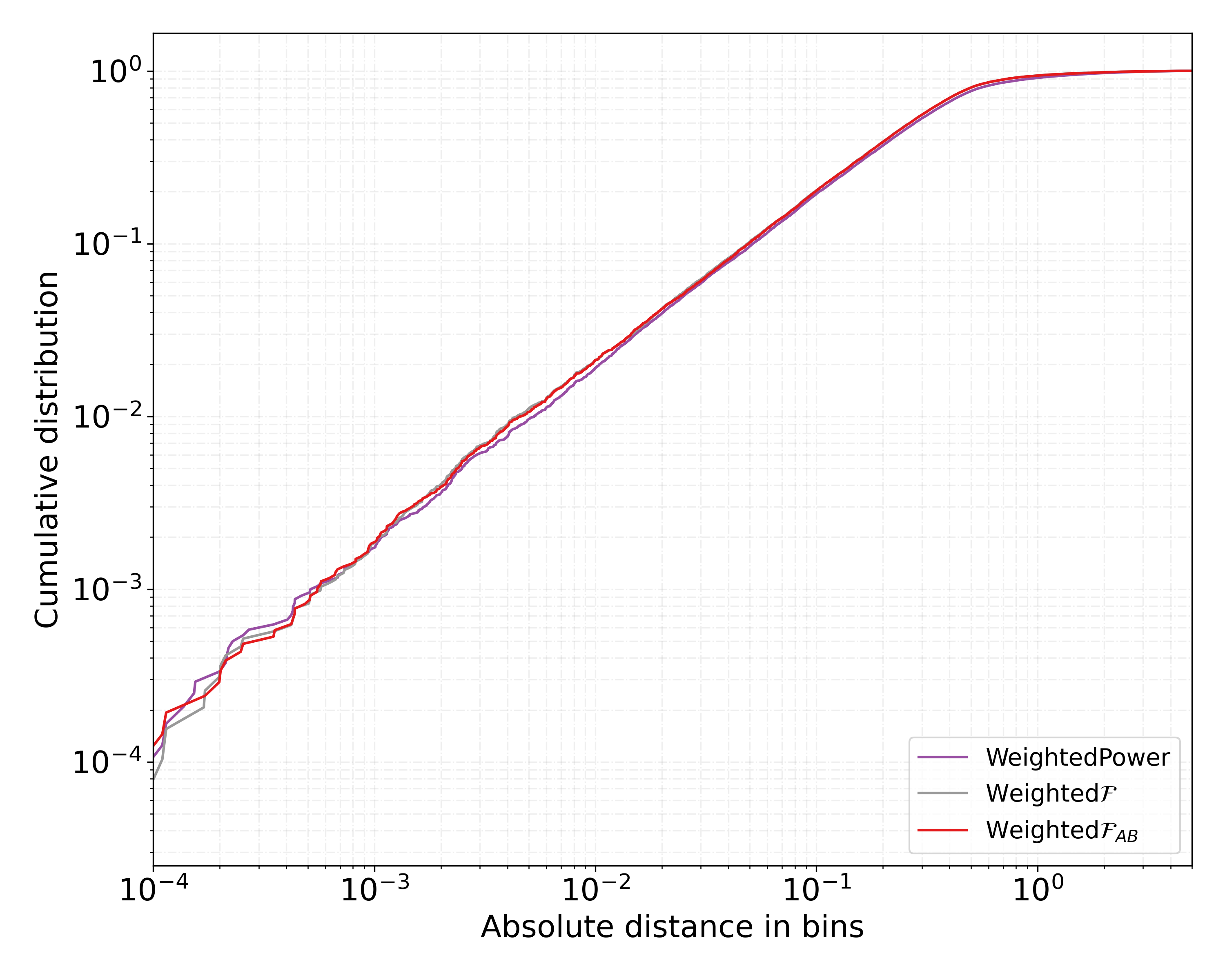}
\caption{Parameter estimation accuracy for different weighted detection statistics.  We plot the
  cumulative distribution of parameter-space offsets between estimate and injected signal, over the
  six parameter-space dimensions searched.  The plot uses \num{1000} injected signals assuming one
  year of data (with equal Gaussian noise floor) from both the H1 and L1 detectors, and a coherent
  segment length of $\Tseg=\SI{900}{\second}$.  }
\label{fig:PE1}
\end{figure}

Next we compare the parameter estimation accuracy of the different weighted detection statistics.
To do this, we select the template with the highest detection statistic, and compare its parameters
with the parameters of the injected signal, if it was detected. The distance between injected signal
and recovered ``loudest'' template is measured in terms of number of grid bins along all six
parameter-space dimensions.  The results are shown in Fig.~\ref{fig:PE1}. It can be seen that the
different detection statistics show a very similar behavior, and that for all of them more than
$90\%$ of the signals are recovered within one bin.

\subsection{Refinement stage}

As discussed in Sec.~\ref{sec:summary-previous-new}, the new \BSHF{} pipeline (like the previous
\BSH{}) consists of two main stages: after the initial search stage, a percentage of templates with
the highest detection statistic are reanalyzed with more accurate $u_{\mathrm{I}k}$ interpolation
expressions (see Sec.~\ref{sec:semi-coher-interp}), and possibly with a finer mismatch.
The previous \BSH{} pipeline also used a more sensitive detection statistic in the second stage,
using the weighted power instead of the weighted Hough number count, while now we always use the
most sensitive statistic (i.e., weighted $\Fab$) in all stages.

The sensitivity of this refinement procedure depends on the percentage of templates that is passed
to the second stage. If it is large enough, at a realistically low false alarm probability the
sensitivity of the search would be determined by the statistic used in the second stage, thus
improving the overall sensitivity.  In order to simulate realistic search conditions, for a
candidate to count as detected in the second stage we also require its first-stage detection
statistic to be higher than the weakest candidate that was passed to the second stage.

\begin{figure}
  \centering
  \includegraphics[width=1.0\columnwidth]{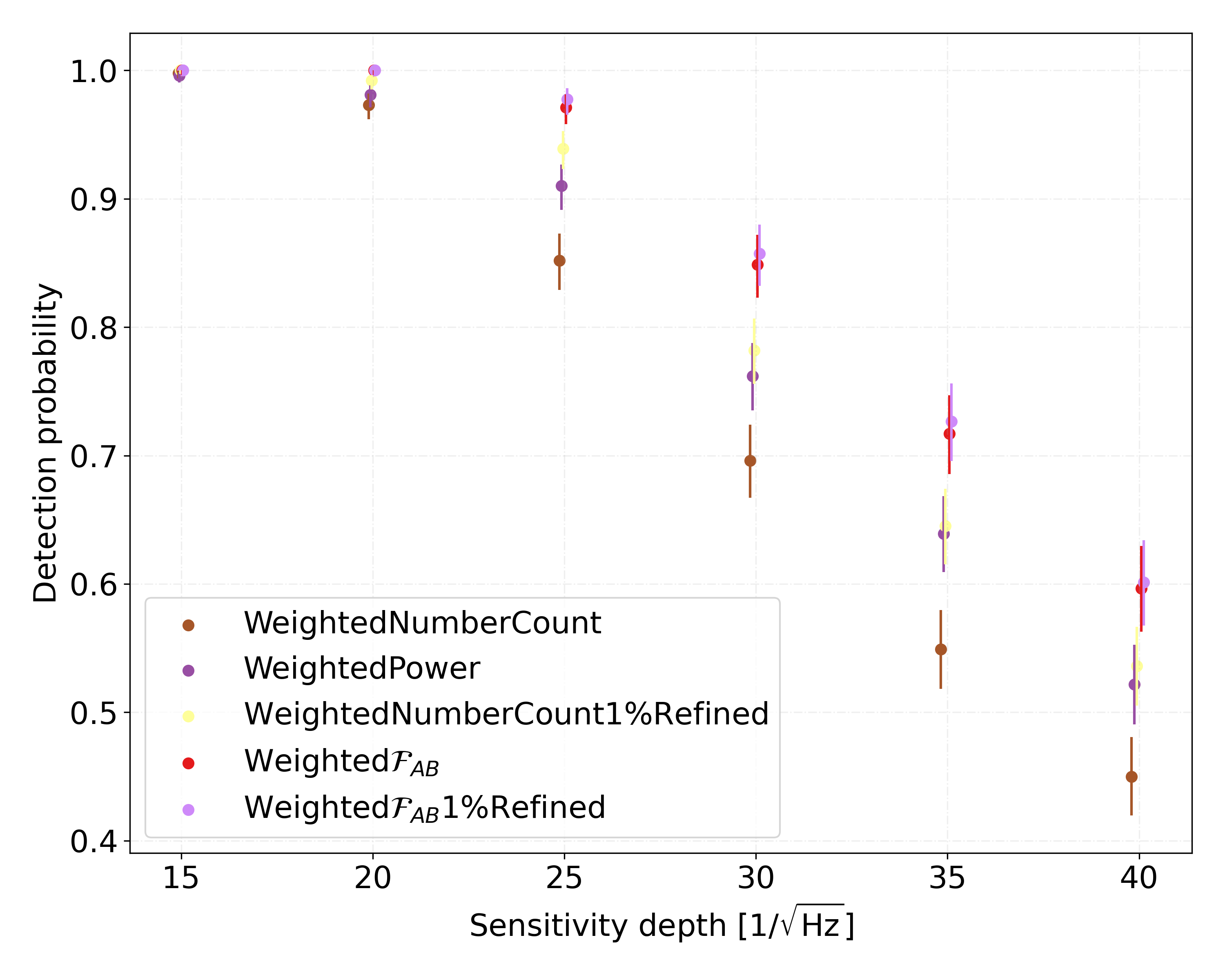}
  \caption{Detection probability as a function of sensitivity depth $\sqrt{S_{\mathrm{n}}}/h_0$ for
    different detection statistics in Gaussian noise.  The plot shows the results for the H1
    and L1 detectors, assuming one year of data at the same noise floor from each of the detectors,
    and a coherent segment length of $\Tseg=\SI{900}{\second}$.  The error bars denote the $95\%$
    binomial confidence interval.  (The points have been slightly displaced along the x-axis to ease
    visibility).
  }
  \label{fig:Sens4}
\end{figure}

Figure~\ref{fig:Sens4} shows the comparison of the weighted number count without a second stage with
the result when the highest $1\%$ of the candidates are passed to the second stage.  We see that the
sensitivity of the weighted number count (with the optimal threshold of 3.2\footnote{Here the
  expectation value of the power statistic is 2, while in \cite{PhysRevD.70.082001} it was 1, which
  explains the factor of two difference.}) is within the uncertainty errors of the weighted
power. This shows that the previous pipeline sensitivity was effectively determined by the
second-stage weighted power statistic.

We further see that the sensitivity of the weighted dominant-response $\Fab$-statistic is slightly
improved in the refinement stage, due to the usage of the more accurate $u_{\mathrm{I}k}$ master
equations of Eqs.~\eqref{eq:21a},~\eqref{eq:21b}.
For more than two detectors, the sensitivities of the demodulated statistics are expected to be even
better, similar to what we saw in the right plot of Fig.\ref{fig:Sens2}.

\subsection{Increasing the coherent time}
\label{sec:SensTc}

One of the advantages of the new pipeline is the ability to extend the coherent segment time $\Tseg$
while maintaining the same mismatch. Using a longer coherent time increases the computational cost,
while also increasing the sensitivity of the search. For a large number of segments, the sensitivity
of a StackSlide search scales as $N\seg^{-1/4}$ (e.g., \cite{Prix_Shaltev_2012}), so at fixed
false-alarm probability and amount of data, the coherent time would need to be increased by a factor
of 16 in order to double the sensitivity of the search.

Here we compare the difference in sensitivity between using a coherent time of
$\Tseg=\SI{900}{\second}$ and $\Tseg=\SI{3600}{\second}$, with fixed maximum mismatch parameters.

For this comparison we add random signals from the binary parameter space region of
$\Porb\in[0.785,0.8]$~days and $\asini\in[0.5,0.6]$~ls.
In this region of the parameter space, several $u_2$ templates are needed in order to maintain the
same coherent mismatch for the coherent time of $\Tseg=\SI{3600}{\second}$, while the searches with
$\Tseg=\SI{900}{\second}$ only need templates in $u_1$.

\begin{figure}
  \centering
  \includegraphics[width=1.0\columnwidth]{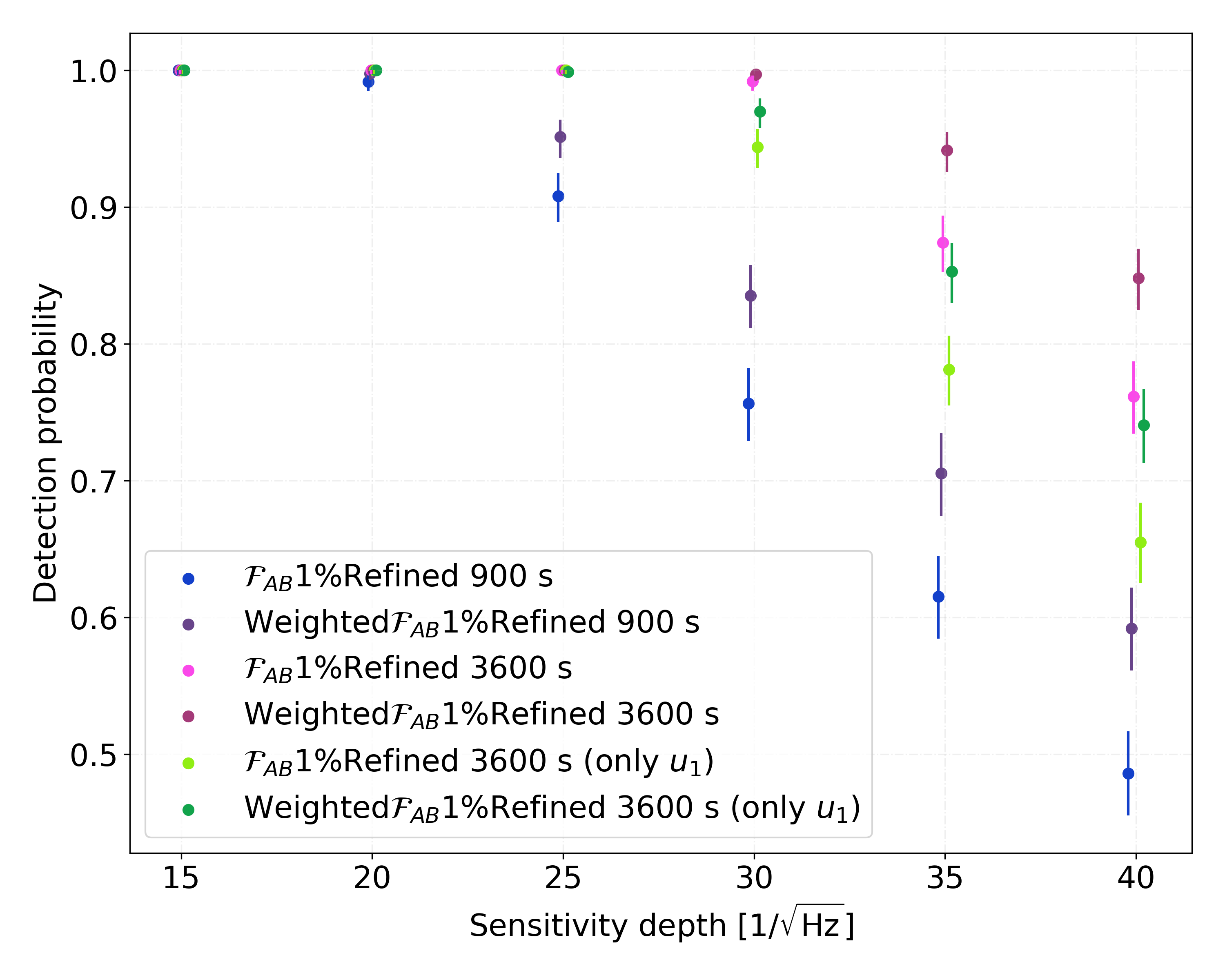}
  \caption{Detection probability as a function of sensitivity depth $\sqrt{S_{\mathrm{n}}}/h_0$ for
    different detection statistics in Gaussian noise, using two different coherent segment lengths,
    namely $\Tseg=\SI{900}{\second}$ and $\Tseg=\SI{3600}{\second}$.
    The plot shows the results for the H1 and L1 detectors, assuming one year of data at the same
    noise floor from each of the detectors. The error bars denote the $95\%$
    binomial confidence interval.  (The points have been slightly displaced along the x-axis to ease
    visibility).
  }
  \label{fig:Sens3}
\end{figure}

Fig.~\ref{fig:Sens3} shows the results by comparing the dominant-response $\Fab$-statistic using
these two coherent times.  The improvement in sensitivity by using a longer coherent time is clear,
both for the weighted and unweighted cases.  For the unweighted detection statistics, the maximal
expected improvement due to the four-times increase in $\Tseg$ is around $\sim1.41$, which agrees
approximately with what is seen in the figure.

An interesting point to observe here is that the relative improvement due to weighting of the
statistics decreases for longer segment length $\Tseg$ (at constant Gaussian-noise floors).
This can be understood from the reduced differences in antenna-pattern responses between segments
which therefore contribute more similar signal power.

We also show in Fig.~\ref{fig:Sens3} a search using the longer $\Tseg=\SI{3600}{\second}$ coherent
segments, but without including $u_2$ templates. We see that the this decreases its sensitivity,
illustrating the need to include $u_2$ templates for this setup, as was expected from the mismatch
estimates in Sec.~\ref{sec:CohMis}.

\section{Conclusions}
\label{sec:conclusions}

In this paper we have presented a new pipeline, \BSHF{}, to search for continuous gravitational
waves from unknown neutron stars in binary systems.  This new pipeline is a descendant of the
previous \BSH{} pipeline, improving several different aspects.
It can cover the same parameter space at a reduced computational cost, and the usage of demodulated
($\mathcal{F}$- and $\Fab$-) statistics allows to use longer coherent segments, which
increases sensitivity and gives more flexibility to the pipeline.
Furthermore, the per-detector data are combined coherently, thereby reducing the computational cost
(by a factor of $\sim\Ndet$) and further improving sensitivity for searches with more than two detectors.

The new pipeline gives explicit control over the mismatch in the coherent stage, allowing one to
perform lower-mismatch searches than before, for example when following up an interesting candidate
or targeting a particularly interesting smaller region of parameter space..
One can now explicitly search over binary orbital parameters such as the eccentricity and argument
of periapse, or higher-order frequency derivatives.
Therefore this pipeline can be used to follow up candidates from different other searches, such as
from an all-sky isolated search that only gives constraints on the ranges of possible binary
parameters to follow-up.

\BSHF{} also has some limitations. The Taylor coordinates used in the coherent stage only allow
searching for orbital periods substantially longer than the coherent segment length, as discussed in
Sec.~\ref{sec:CohMis}. This limits the possibilities to search over the shortest orbital periods
with longer coherent times. The corresponding RAM requirements limit the number of Taylor
coordinates that can be used, which also limits the maximum coherent time that can be used in
certain regions of the parameter space.

However, a number of future improvements to this pipeline can be envisaged.  For example, the usage
of a non-zero threshold on the coherent statistics before summing, or including only a subset of
segments in the first stage depending on their sensitivity.  Further enhancements could be achieved
by reducing the second-stage mismatch by refining the search grid. It would also be interesting to
develop an explicit optimization algorithm for the optimal search setup parameters given a certain
amount of data and computational budget, similar to what was done in
\cite{Prix_Shaltev_2012,ming_optimally_2018}.  We also plan to explore the usage of further
demodulated detection statistics, in particular the line-robust statistics of
\cite{2014PhRvD..89f4023K}, which should help mitigate against non-Gaussian artifacts in the data.

\begin{acknowledgments}
This work has utilized the ATLAS computing cluster at the MPI for Gravitational Physics Hannover.

This project has received funding from the European Union's Horizon 2020 research and innovation programme under the Marie Sklodowska-Curie grant agreement number 101029058.
\end{acknowledgments}

\appendix
\section{Detector-frame Taylor coordinates}
\label{sec:appendix}

In Sec.~\ref{sec:phase} we introduced short-segment SSB Taylor coordinates $\{u_k\}$, defined by
the Taylor expansion of the signal phase in the SSB frame around each segment mid-point.
Here we show how similar \emph{detector-frame} Taylor coordinates $\{u_k^X\}$ can be expressed for
the phase evolution in the frame of detector $X$, which additionally includes the phase modulation
due to the detector motion.

In complete analogy to Eq.~\eqref{eq:3} we can write the Taylor-expansion in detector arrival time
$t$ around a mid-time $\tmid$:
\begin{equation}
  \label{eq:ap1}
  \phi^X(t) = \phi_0 + 2\pi \sum_{k=1}^\kmax \frac{u^X_k}{k!}\, (t - \tmid)^k\,,
\end{equation}
with the detector-frame Taylor coordinates $\{u_k^X\}_{k=1}^\kmax$ defined as:
\begin{equation}
  \label{eq:ap2}
  u^X_k \equiv \frac{1}{2\pi}\, \left.\frac{d^k\phi^X}{d t^k}\right|_{\tmid}\,,
\end{equation}
in terms of the timing relation between source-frame (emission) time $\tau$ and detector-frame
(arrival) time $t$, obtained by combining Eqs.~\eqref{eq:sourceSSB} and \eqref{eq:2}:
\begin{equation}
  \label{eq:7}
  \tau^X(t) = \tau(\tSSB^X(t)) = t + \vrr^X(t)\cdot\vn - R(\tau^X).
\end{equation}
The expressions Eq.~\eqref{eq:6} are formally identical, with the SSB derivatives $\tau^{(k)}$
replaced by detector-time derivatives $\tau^{X(k)}\equiv d^k\tau^X/dt^k$, which are found as
\begin{equation}
  \label{eq:15}
    \begin{aligned}
      \dot{\tau}^X &= \left[1+R'\right]^{-1} (1 + \dot{\vrr}^X\cdot\vn) ,\\
      \ddot{\tau}^X &= \left[1+R'\right]^{-1}\left(\ddot{\vrr}^X\cdot\vn - R''\,\dot{\tau}^X{}^2\right),\\
      \dddot{\tau}^X &= \left[1+R'\right]^{-1}\left(\dddot{\vrr}^X\cdot\vn -R''\,\dot{\tau}^X{}^3 - 3R''\,\ddot{\tau}^X\dot{\tau}^X\right),\\
      \vdots
  \end{aligned}
\end{equation}
which generalizes Eq.~\eqref{eq:18}.

We can write the first two orders explicitly as
\begin{equation}
  \label{eq:17}
  \begin{aligned}
    \frac{u^X_1}{f\mmid} &= \frac{1 + \vv\mmid^X\cdot\vn}{1 + R'\mmid},\\
    u^X_2 &= \frac{f\mmid}{1 + R'\mmid}\left(\va\mmid^X\cdot\vn -
      R''\mmid\left(\frac{u^X_1}{f\mmid}\right)^2\right) + f'\mmid\,\left(\frac{u^X_1}{f\mmid}\right)^2,
  \end{aligned}
\end{equation}
with the definitions of Eqs.~\eqref{eq:13} and \eqref{eq:16} (for small eccentricity and a single
spindown), and $\vv^X\mmid\equiv \dot{\vrr}^X(\tmid)$, $\va^X\mmid \equiv \ddot{\vrr}^X(\tmid)$ and
detector velocity and acceleration at the segment mid-time $\tmid$, respectively.
Again, these coordinates have units of Hz and \si{Hz^2} respectively, but now also depend on the
detector $X$ and on the sky position $\vn$ of the signal.

Assuming no spindown $f_1=0$, a circular orbit ($\ecc=0$), and a non-relativistic orbital velocity
$\asini\Om \ll 1$, Eq.~\eqref{eq:17} yields an approximate equation for the frequency-time pattern,
namely
\begin{equation}
  \label{eq:approx}
  \begin{aligned}
    u_1^X &\approx f_0 \,( 1 + \vv^X\mmid\cdot\vn) \, ( 1 - \asini\Om\cos\Psi\mmid ) \\
    & \approx f_0 + f_0 \,\vv^X\mmid\cdot\vn - f_0\, \asini\Om\cos\Psi\mmid,
  \end{aligned}
\end{equation}
which is the same as Eq.~(15) of \cite{2019PhRvD..99l4019C}, evaluated at segment mid-time $\tmid$.

\bibliography{biblio}

\end{document}

%% file: git_tag.tex
\newcommand{\commitDATE}{2022-08-05 11:54:07 +0200}
\newcommand{\commitID}{commitID: 37f10b7bec88299dc29219f40930f0f0928db8c6}
\newcommand{\commitIDshort}{commitID: 37f10b7}
\newcommand{\commitSTATUS}{CLEAN}